\documentclass[prd,superscriptaddress,groupedaddress,nofootinbib,12pt]{revtex4-2}
\usepackage{amsmath}
\usepackage[colorlinks=true]{hyperref}
\usepackage{upgreek}
\usepackage{amsxtra}
\usepackage{tikz}
\usepackage{stix}
\usepackage{pgfplots}
\usepackage{subcaption}
\usepackage{wrapfig}
\usepackage{ulem}
\usetikzlibrary{decorations.markings}
\usetikzlibrary{arrows.meta}

\renewcommand{\i}{\mathrm{i}}      
\renewcommand{\d}{\mathrm{d}}      
\newcommand{\p}{\partial}          
\newcommand{\h}{\hbar}             
\newcommand{\e}{\mathrm{e}}        
\renewcommand{\pi}{\uppi}          

\begin{document}
	
	\title{Dawn and Twilight Time in Quantum Tunneling}
	\author{Tinglong Feng}
	\email{t.feng@students.uu.nl}
	\author{Jesse Moes}%
	\author{Tomislav Prokopec}
	\email{t.prokopec@uu.nl}
	\affiliation{Institute for Theoretical Physics, Utrecht University, Princetonplein 5, 3584 CC, Utrecht, Netherlands}%
	
\begin{abstract}
	Metastable decay exhibits a familiar exponential regime bracketed by early time deviations and late-time power-law tails. We adopt the real-time, flux-based definition of the decay rate in the spirit of Andreassen et al. direct method and present a complete analysis of one-dimensional quantum mechanical resonance models. We show that the kernel admits a universal pole–plus–branch decomposition and use it to define two computable time scales: a {\it dawn time} $t_\circlebottomhalfblack$,   when a single resonant contribution starts dominating and exponential decay sets in, and a {\it twilight time} $t_\circletophalfblack$, when the branch-cut tail overtakes exponential decay.  The latter can be expressed in closed form {\t via} the Lambert $W$ function, making its parametric dependence manifest without fitting. For square, modified square, and P\"oschl--Teller barriers we obtain simple thick-barrier formulas, clarify the relation $\Gamma T = T_{\rm trans}$ between the decay rate $\Gamma$, oscillation period $T$, and transmission probability $T_{\rm trans}$, and indicate how our spectral picture can be naturally extended to quantum field theoretic vacuum decay.
\end{abstract}

	\maketitle
	
	
\section*{Introduction}
Metastable quantum states are ubiquitous, from nuclear $\alpha$ decay and cold-atom tunneling to vacuum decay in the early Universe. The statistical theory of 
bubble nucleation was originally formulated by Langer~\cite{Langer:1967ax,Langer:1969bc}. Langer's 
theory was subsequently extended to quantum field theory by
Coleman and 
Callan~\cite{Coleman:1977py,Callan:1977pt}, who computed false-vacuum decay by analytically continuing the real-time action to imaginary (Euclidean) time, finding bounce solutions and evaluating fluctuation determinants, thereby fixing both the decay rate and its parametric scaling. 
The Euclidean framework has been extensively developed and generalized, to include thermal and other effects~\cite{Affleck:1980ac,kleinert2009path,Muller-Kirsten:2012wla,Zinn-Justin:2002ecy,Marino:2015yie,Weinberg_2012}, and 
in recent years there has been substantial progress in precision computations of false-vacuum decay rates~\cite{Ai:2019fri, Ai:2023yce, Matteini:2024xvg, Batini:2023zpi, Pirvu:2023plk, Pirvu:2024nbe, Kierkla:2025qyz, Ai:2024ser, Carosi:2024lop}. 
However, the Euclidean framework obscures real-time dynamics, as it requires analytic continuation of the potential.

In parallel, real-time, flux-based approaches, most notably Schwartz's direct
method~\cite{Andreassen:2016cff,Schwartz:DirectMethod}, define decay through probability current exiting a metastable region, retaining scattering intuition without analytic continuation. 
Building on this real-time perspective for one-dimensional quantum-mechanical resonance models, we identify two in practice computable time scales: {\it the dawn time}, when a single resonant pole begins dominating (which depends on the initial-state), and {\it the twilight time}, when exponential decay gives way to the universal power-law tail (which is insensitive to the initial-state). We provide (i) a compact flux formulation using a pole-branch decomposition of the real-time kernel, (ii) closed-form expressions for both time scales with the twilight time expressed via a {\it Lambert $W$ function}.
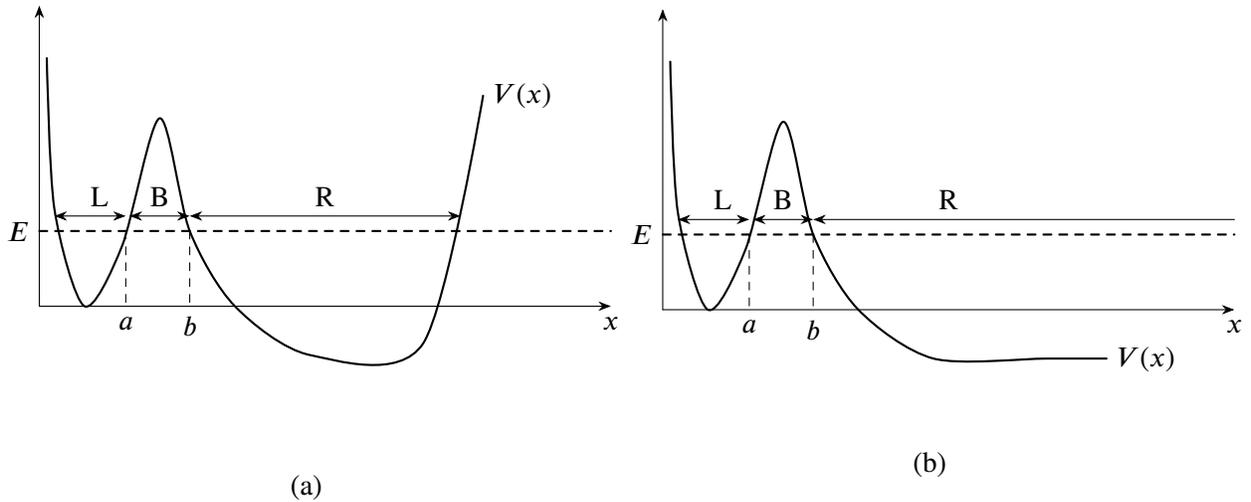
\begin{figure}[h]
	\begin{subfigure}[t]{0.5\textwidth}
		\centering
			\begin{tikzpicture}[>=Stealth, line cap=round, line join=round, scale=1]
			\draw[->] (-0.6,0) -- (7.0,0) node[below] {$x$};
			\draw[->] (-0.6,0) -- (-0.6,4);
			\draw [dashed,thick](-0.6,1) -- (7,1) ;
			\draw [dashed](0.55,1) -- (0.55,0) ;
			\draw [dashed](1.4,1) -- (1.4,0) ;
			\draw[thick]
			plot[smooth]
			coordinates{
				(-0.50, 3.30)
				(-0.40, 1.30)
				( 0.00, 0.00)   
				( 0.5, 0.8)
				( 1.0, 2.5)    
				( 1.4, 1)
				( 2, 0)    
				( 3,-0.65)
				( 4.5,-0.5)
				( 5.30, 2.80)
			};
			
			\node[below] at (0.55,0) {$a$};
			\node[below] at (1.4,0) {$b$};

			\draw[<->] (-0.40,1.2) -- (0.55,1.2);
			\node[above] at (0.2,1.2) {L};

			\draw[<->] (1.4,1.2) -- (5,1.2);
			\node[above] at (3.2,1.2) {R};
			\node[above] at (1,1.2) {B};
			\draw[<->] (0.6,1.2) -- (1.4,1.2);
			\node[right] at (5.30,2.8) {$V(x)$};
			\node[left] at (-0.6,1) {$E$};
		\end{tikzpicture}
		\label{(a)}
		\caption{}
	\end{subfigure}
~
\begin{subfigure}[t]{0.5\textwidth}
	\centering
		\begin{tikzpicture}[>=Stealth, line cap=round, line join=round, scale=1]
		\draw[->] (-0.6,0) -- (7.0,0) node[below] {$x$};
		\draw[->] (-0.6,0) -- (-0.6,4);
		\draw [dashed,thick](-0.6,1) -- (7,1) ;
		\draw [dashed](0.55,1) -- (0.55,0) ;
		\draw [dashed](1.4,1) -- (1.4,0) ;
		\draw[thick]
		plot[smooth]
		coordinates{
			(-0.50, 3.30)
			(-0.40, 1.30)
			( 0.00, 0.00)   
			( 0.5, 0.8)
			( 1.0, 2.5)    
			( 1.4, 1)
			( 2, 0)    
			( 3,-0.65)
			( 4.5,-0.65)
			( 5.30, -0.65)
		};
		
		\node[below] at (0.55,0) {$a$};
		\node[below] at (1.4,0) {$b$};

		\draw[<->] (-0.40,1.2) -- (0.55,1.2);
		\node[above] at (0.2,1.2) {L};

		\draw[<-] (1.4,1.2) -- (7,1.2);
		\node[above] at (3.2,1.2) {R};
		\node[above] at (1,1.2) {B};
		\draw[<->] (0.6,1.2) -- (1.4,1.2);
		\node[right] at (5.30,-0.65) {$V(x)$};
		\node[left] at (-0.6,1) {$E$};
	\end{tikzpicture}
	\label{(b)}
\vskip -0.3cm
	\caption{}
\end{subfigure}
	\caption{Two examples of a potential $V(x)$ with a metastable region L, a destination region R, and a barrier B. For case (a), in which R is finite, the wave function initially localised in L will not completely decay, as eventually a strong back-reaction will kick in, such that the probability will oscillate between regions L and R. For case (b), in which R extends to $+\infty$, no significant back-reaction will occurs, and an initial state localized in L will continue decaying into R until it decays completely.
    \hskip 2cm}
	\label{example potential}
\end{figure}
First, we must give a proper definition of tunneling rates. Consider a potential $V(x)$ with a barrier shown in figure~\ref{example potential}. We label the
local minimum inside the left region (L) by $a$ and the turning point by $b$, which is defined by $V(a)=V(b)=E$, $E$ is the energy of the particle initially trapped in L. We want to calculate the probability for the wavefunction to still be in the false vacuum at a certain time, and relate the decay of this probability to a tunneling rate.

\noindent The rate can be obtained from quantum–mechanical flux conservation, {\it i.e.} the continuity equation,
\begin{equation}
	\partial_t |\Psi|^2 + \nabla \!\cdot\! \mathbf{j} = 0
\,,\quad
\end{equation}
where $ |\Psi|^2$ is the probability density, 
and the probability density current is defined as
\begin{equation}
	\mathbf{j}
	=\; \frac{\hbar}{2 m i}\!\left\{\Psi^{\dagger} \nabla\Psi
	- \Psi\, \nabla \Psi^{\dagger}\right\}
\,,\quad
\end{equation}
which follows from the Schr\"odinger equation. If one assumes outgoing boundary
conditions, an intuitive definition for the decay rate is the
fraction of the probability flowing out of the left legion,
\begin{equation}
	\Gamma \;=\; -\,\frac{ \int_{\partial V} \d\Sigma^{\,i}\, j_i}
	{\displaystyle \mathbb{P}\bigl(t,x\in \text{V}\bigr)}\,,
    \qquad 	
    {\displaystyle \mathbb{P}\bigl(t,x\in \text{V}\bigr)}
    =\int_{\mathrm{V}}|\Psi(t,x)|^2\d x
\,,\quad
\end{equation}
where $\d\Sigma^{\,i}$ denotes the oriented surface element on $S=\partial V$. Adapting to the case of a particle trapped in the one-dimensional potential considered here, we have $V\rightarrow L$ and $ \int_{\partial V} \d\Sigma^{\,i}\, j_i \rightarrow j(x\!=\!a)$.
If the
energy eigenfunctions form a complete set, an arbitrary initial wavefunction can be
expanded in these states. Inserting a single energy eigenfunction gives the tunneling
rate associated with that specific mode.

\begin{wrapfigure}[18]{l}{0.36\textwidth}
	\centering
	\begin{tikzpicture}[scale=1, >=Stealth]
		\draw[->] (-0.2,0) -- (4.5,0) node[right] {$\mathrm{Re} E$};
		\draw[->] (0,-4) -- (0,4.2) node[above] {$\mathrm{Im} E$};
		\draw[red,very thick,->] (0,-0.1) -- (4,-0.1);
		\draw[red,very thick,->,dashed] (0,0.1) -- (4,0.1) node[below right] {$\infty$};
		\draw[red,very thick,->] (0,-4) -- (0,-0.1);
		\draw[red,very thick,->,dashed] (0,4) -- (0,0.1);
		\draw[red,very thick,->,] (4,-0.1) arc[start angle=0,end angle=-90,radius=4];
		\draw[red,very thick,->,dashed] (4,0.1) arc[start angle=0,end angle=90,radius=4];
		\coordinate (pole0) at (1.5,-1.0);
		\coordinate (pole0*) at (1.5,1.0);
		\coordinate (pole1) at (2.5,-2);
		\coordinate (pole1*) at (2.5,2);
		\fill[red] (pole0) circle (2pt);
		\fill[red] (pole0*) circle (2pt);
		\fill[cyan!70!blue] (pole1) circle (2pt);
		\fill[cyan!70!blue] (pole1*) circle (2pt);
		\node[above] at (pole0) {$E_0-\tfrac{\i\h}{2}\Gamma_0$};
		\node[above] at (pole1) {$E_1-\tfrac{\i\h}{2}\Gamma_1$};
		\node[above] at (pole0*) {$E_0+\tfrac{\i\h}{2}\Gamma_0$};
		\node[above] at (pole1*) {$E_1+\tfrac{\i\h}{2}\Gamma_1$};
		\node at (3.0,-2.5) {$\mathcal{C}$};
		\node[right] at (0,-2.5) {$\mathcal{C}$};
		\node[below] at (3.0,-0.25) {$\mathcal{C}$};
		\node[above] at (3.0,0.25) {$\mathcal{C'}$};
		\node at (2,3.3) {$\mathcal{C'}$};
		\node[right] at (0,2.5) {$\mathcal{C'}$};
	\end{tikzpicture}
	\caption{Contour $\mathcal{C}$ and the poles of the kernel in the complex energy plane. The arc at infinity does not contribute because the integrand decays exponentially in this limit.}
	\label{contour0}
\end{wrapfigure}
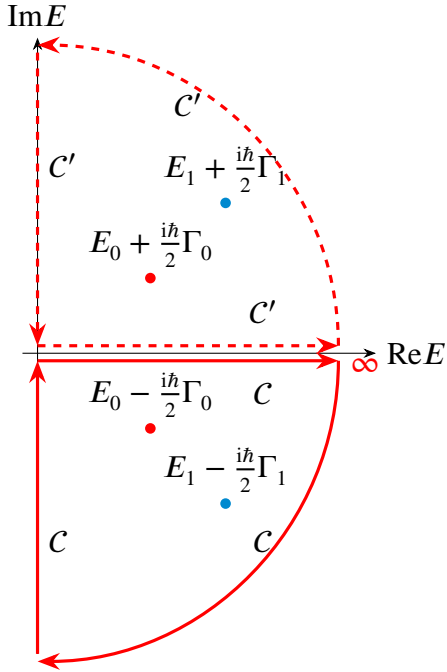

\noindent In order to obtain the decay rate, we must first compute the kernel $K(t,x;0,y)$~\footnote{The quantum-mechanical kernel is defined as the projection of the evolution operator $\hat U(t;0)$ on the position space eigenstates $|x\rangle$, $K(t,x;0,y)=\langle x|\hat U(t;0)|y\rangle$. It is sometimes called the propagator, however we prefer to reserve this notion for the time-ordered two-point function.} with $x,y\in \text{L}$. Using kernel allows to seperate what comes from initial state, and what from the nature of the system. In particular, \textit{dawn time} depends on the initial state, \textit{twilight time} can be calculated from the kernel, and therefore it is independent on the initial state.  In path-integral quantization we have~\cite{Feynman:100771},
 \begin{align}
	K(t,x;0,y)&=\int_{0}^{\infty} \frac{\d p}{2\pi\h} \,
	\phi_{p}^{*}(y)\,\phi_{p}(x)\,
	\e^{-\i\frac{p^{2}}{2m\h}t}
\nonumber\\
	&=\int_{0}^{\infty} \frac{\d E}{2\pi\h} \sqrt{\frac{m}{2E}} \,
	\phi_{p}^{*}(y)\,\phi_{p}(x)\,
	\e^{-\frac{\i}{\h} Et}
\,,\quad
\label{kernel in terms of eigenfunctions}
\end{align}
where $\phi_{p}(x)$ are the eigenfunctions of the Schr\"odinger equation $E\phi_{p}(x) = \hat H(x)\phi_{p}(x)$, with eigenvalue $p>0$ and $E(p)=p^2/(2m)$.
We can evaluate the kernel 
in~(\ref{kernel in terms of eigenfunctions}) by using contour integration. For a forward evolution in time ($t>0$) the suitable contour is $\mathcal{C}$ in figure~\ref{contour0}, in which the arc at infinity does not contribute.~\footnote{For a backward evolution in time ($t<0$) the contour $\mathcal{C}'$ in figure~\ref{contour0} should be closed above the real axis. In which case the imaginary parts of the energy poles have the opposite sign, 
leading to damping. This explains how the quantum mechanical evolution, which originates from the Schr\"odinger differential equation which is symmetric under $t\rightarrow -t$, can violate time reversal symmetry.}
With this in mind, the integral in~(\ref{kernel in terms of eigenfunctions}) can be expressed as,

\begin{equation}
	K(t,x;0,y)=f(t,x,y)+g(t,x,y)
\,,\quad
\label{kernel = f + g}
\end{equation}
where
\begin{equation}
	f(t,x,y)=\oint_{\mathcal{C}}  \frac{\d E}{2\pi\h} \sqrt{\frac{m}{2E}}
	\phi_{p}^{*}(y)\,\phi_{p}(x)\,
	\e^{-\frac{\i}{\h} Et}
\,,\quad
\label{function f}
\end{equation}
and
\begin{align}
	g(t,x,y) &= \int_{0}^{-\i\infty}  \frac{\d E}{2\pi\h} \sqrt{\frac{m}{2E}}
	\phi_{p}^{*}(y)\,\phi_{p}(x)\,
	\e^{-\frac{\i}{\h} Et}\notag \\[8pt]
	&= -\,\i \int_{0}^{\infty} \frac{\d\mathcal{E}}{2\pi\h}  \sqrt{\frac{m}{-\i\,2\mathcal{E}}}
	\phi_{p}^{*}(y)\,\phi_{p}(x)\,
	\e^{-\mathcal{E}t/\h}
\,.\quad
\end{align}
%
%
At early times $g(t,x,y)$ can be calculated by expanding the kernel in powers of $t$. This expansion is valid as long as $\frac{t\h}{2m}\biggl|\frac{\frac{\p^2}{{\p x}^2}\psi(0,x)}{\psi(0,x)}\biggr|\ll1$. This means that late (asymptotic) time expansion is valid when,
\begin{equation}
	\frac{t\h}{2m}\Biggl|\frac{\frac{\p^2}{{\p x}^2}\psi(0,x)}{\psi(0,x)}\Biggr|\gg 1
\,.\quad
\end{equation}

%
 The function $f(t,x,y)$ in~(\ref{function f}) is evaluated by the residue theorem. For the three concrete potentials in the Appendices, the kernel scales like,
\begin{align}
	K(t,x;0,y)
	&\approx 
	G(E_0,x,y)\,\left(\frac{E_{0}\, t}{\h}\right)^{-\gamma}
\nonumber\\
    &\hskip 0.4cm
    + \sum_{n}F(E_n,x,y)\exp\!\left\{-\,\frac{\Gamma_{n}}{2}\, t\right\}\label{late time kernel}
	\\[4pt]
	&\hskip -1.cm
    \approx
	G(E_{0},x,y)\,\left(\frac{E_{0}\, t}{\h}\right)^{-\gamma}
	+ F(E_{0},x,y)\exp\!\left\{\!-\,\frac{\Gamma_{0}}{2}\, t\right\},
\end{align}
where the last approximation holds for sufficiently late times. For the square barrier potential (Appendix~\ref{suqre barrier}), $\gamma=3/2$; for the modified square barrier potential (Appendix	~\ref{Modified square barrier}), $\gamma=2$; and for the P\"oschl-Teller potential (Appendix	~\ref{P\"oschl-Teller potential}), $\gamma=1/2$. 

\section*{Twilight time}

The mnemonic term {\it twilight time} $t_\circletophalfblack$ is chosen for the time when the power-law tail starts to dominate over the exponentially decaying regime, and it is independent
of the initial state. To investigate the time regime where the first-order pole contribution
dominates, consider first the equation,
\begin{equation}
	\left(\frac{E_{0}t_\circletophalfblack}{\h}\right)^{-2\gamma}={\Bigg|\frac{F}{G}\Bigg|}^2\exp(-\Gamma_{0}t_\circletophalfblack)
\,,\quad
\label{twilight}
\end{equation}
where $|F|=|F(E_0,x,y)|,|G|=|G(E_0,x,y)|$. Note that there are two intersection points, one at an early time and one at a late time, the later one corresponding to $t_\circletophalfblack$. Let us first rewrite \eqref{twilight} as,
\begin{equation}
	t_\circletophalfblack=\frac{\h}{E_{0}}\Bigg|\frac{G}{F}\Bigg|^{1/\gamma}\!
	\exp\!\left(\frac{\Gamma_{0}}{2\gamma}\,
    t_\circletophalfblack\right)
\,.\quad
\end{equation}
Defining $\tau=\tfrac{\Gamma_{0}}{2\gamma}t_\circletophalfblack$ and multiplying by $-\frac{\Gamma_{0}}{2\gamma}$, one obtains
\begin{equation}
	-\tau\,\exp(-\tau)=-\,\frac{\Gamma_{0}}{2\gamma}\,\frac{\h}{E_{0}}
	\Bigg|\frac{G}{F}\Bigg|^{1/\gamma}
\,.\quad
\end{equation}
Compare this to the defining transcendental equation of the Lambert $W$ function,
%
	$W_{k}(z)\,\exp\{W_{k}(z)\}=z$,
%
we then have,
\begin{equation}
	\tau
	= -W_{k}\!\left(
	-\,\frac{\h}{E_{0}}
	\Bigg|\frac{G}{F}\Bigg|^{1/\gamma}
	\frac{\Gamma_{0}}{2\gamma}
	\right),
	\qquad (k=0,-1)
\,,\quad
\end{equation}
where the branches $k=0,-1$ follow from the requirement that $\tau$ is real.
According to the properties of the Lambert $W$ function~\cite{LambertW}, 
the intersection point at a later time is,
\begin{equation}
	t_\circletophalfblack
	= -\,\frac{2\gamma}{\Gamma_{0}}\,
	W_{-1}\!\left(
	-\,\frac{\h}{E_{0}}
\Bigg|\frac{G}{F}\Bigg|^{1/\gamma}
	\frac{\Gamma_{0}}{2\gamma}
	\right)
\,,\quad
\end{equation}
which is the twilight time. The earlier intersection point,
\begin{equation}
	t
	= -\,\frac{2\gamma}{\Gamma_{0}}\,
	W_{0}\!\left(
	-\,\frac{\h}{E_{0}}
	\Bigg|\frac{G}{F}\Bigg|^{\!{1/\gamma}}\!
	\frac{\Gamma_{0}}{2\gamma}
	\right)
\label{W0}
\,,\quad
\end{equation}
will be used in the discussion of dawn time.
\begin{figure}[h]
	\begin{subfigure}[t]{0.5\textwidth}
		\begin{tikzpicture}[scale=0.83]
			\begin{axis}[
				width=10cm, height=7cm,
				xlabel={$v=\kappa (b-a)/\h$}, ylabel={$\Gamma_{0}t_\circletophalfblack(v)$},
				legend style={draw=none, at={(0.02,0.98)}, anchor=north west}
				]
				\addplot+[very thick,mark=none,red] table [col sep=comma] {Gamma0t2_vs_kappa0d_SB_6.csv};
				\addlegendentry{$\omega_{\mathrm{A}}=6$}
				\addplot+[very thick,mark=none,orange] table [col sep=comma] {Gamma0t2_vs_kappa0d_SB_8.csv};
				\addlegendentry{$\omega_{\mathrm{A}}=8$}
				\addplot+[very thick,mark=none,blue] table [col sep=comma] {Gamma0t2_vs_kappa0d_SB_10.csv};
				\addlegendentry{$\omega_{\mathrm{A}}=10$}
			\end{axis}
		\end{tikzpicture}
		\caption{Twilight time plot for Appendix~\ref{suqre barrier} with varying dimensionless parameter $\omega_{\mathrm{A}}=a\sqrt{2mV_0}/\h$.}
		\label{twilightA}
\end{subfigure}
~
\begin{subfigure}[t]{0.5\textwidth}
	\centering
	\begin{tikzpicture}[scale=0.83]
		\begin{axis}[
			width=10cm, height=7cm,
			xlabel={$v=\kappa (b-a)/\h$}, ylabel={$\Gamma_{0}t_\circletophalfblack(v)$},
			legend style={draw=none, at={(0.02,0.98)}, anchor=north west}
			]
			\addplot+[very thick,mark=none,red] table [col sep=comma] {Gamma0t2_vs_kappa0d_mSB_6.csv};
			\addlegendentry{$\omega_{\mathrm{B}}=6$}
			\addplot+[very thick,mark=none,orange] table [col sep=comma] {Gamma0t2_vs_kappa0d_mSB_8.csv};
			\addlegendentry{$\omega_{\mathrm{B}}=8$}
			\addplot+[very thick,mark=none,blue] table [col sep=comma] {Gamma0t2_vs_kappa0d_mSB_10.csv};
			\addlegendentry{$\omega_{\mathrm{B}}=10$}
		\end{axis}
	\end{tikzpicture}
	\caption{Twilight time plot Appendix~\ref{Modified square barrier} with varying dimensionless parameter $\omega_{\mathrm{B}}=a\sqrt{2m(V_0-V_1)}/\h$.}
\label{twilightB}
\end{subfigure}
\caption{Twilight time plot for the square barrier and modified square barrier.}
\label{twilightAB}
\end{figure}
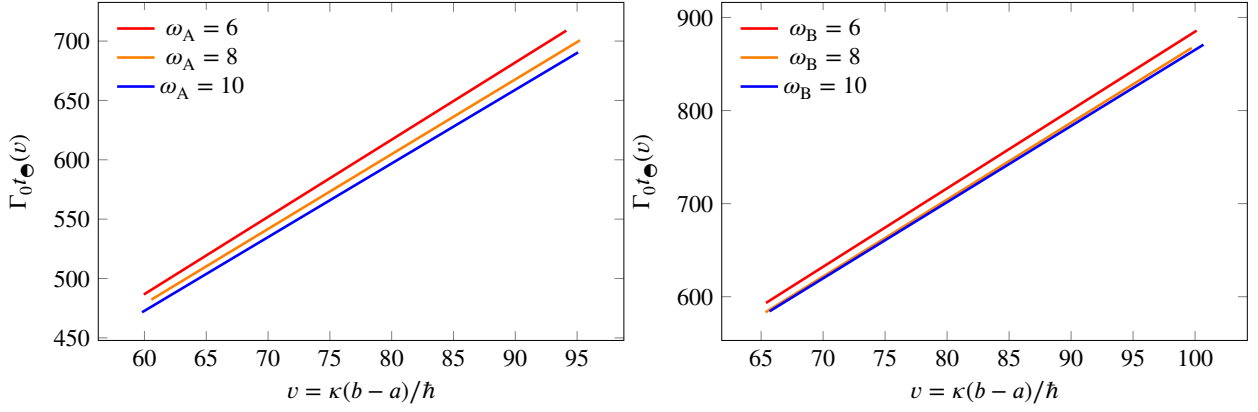
\begin{figure}[h]
	\centering
	\begin{tikzpicture}
		\begin{axis}[
			width=10cm, height=7cm,
			legend style={draw=none, at={(0.02,0.98)}, anchor=north west},
			xlabel={$v=\pi\kappa/(\alpha\h)$}, ylabel={$\Gamma_{0}t_\circletophalfblack(v)$},
			]
			\addplot+[very thick,mark=none,red] table [col sep=comma] {Gamma0t_vs_piKappa_over_alpha_40.csv};
			\addlegendentry{$\omega_{\mathrm{C}}=40$}
			\addplot+[very thick,mark=none,orange] table [col sep=comma] {Gamma0t_vs_piKappa_over_alpha_100.csv};
			\addlegendentry{$\omega_{\mathrm{C}}=100$}
			\addplot+[very thick,mark=none,blue] table [col sep=comma] {Gamma0t_vs_piKappa_over_alpha_200.csv};
			\addlegendentry{$\omega_{\mathrm{C}}=200$}
		\end{axis}
	\end{tikzpicture}
	\caption{Twilight time plot for P\"oschl-Teller potential in Appendix \ref{P\"oschl-Teller potential} with varying dimensionless parameter $\omega_{\mathrm{C}}=b\sqrt{2mU_0}/\h$.}\label{twilight2}
\end{figure}
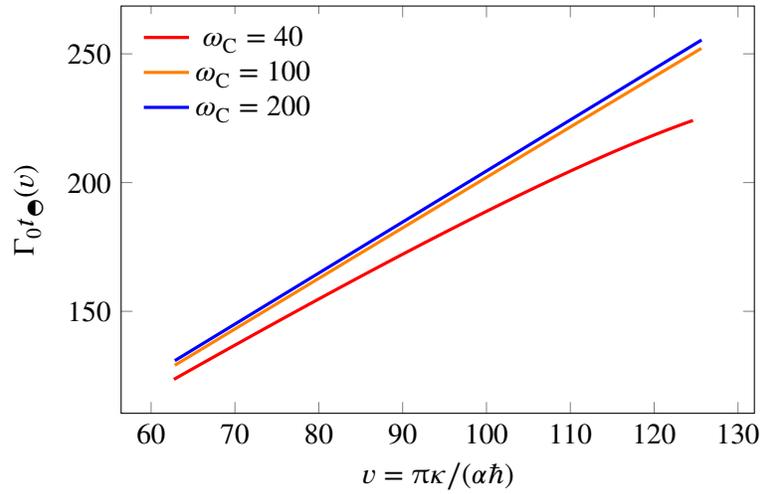
We can now calculate some examples. In the thick-barrier case $\bigl(\kappa(b-a)/\h\gg1$, $\kappa=\sqrt{2mV_0-p_0^2}$, $p_0$ is obtained from \eqref{k_0AB}$\bigr)$, for the square barrier potential in Appendix \ref{suqre barrier} and the modified square barrier potential in Appendix~\ref{Modified square barrier}, the following approximate formula is obtained,
\begin{equation}
	t_\circletophalfblack\approx\frac{4\gamma\kappa(b-a)}{\Gamma_{0}\h}
\,,\quad
\label{twilight time: estimate}
\end{equation}
where $\gamma=3/2$ for the square barrier and $\gamma=2$ for the modified square barrier. 
Figure~\ref{twilightA} shows a plot of $\Gamma_{0}t_\circletophalfblack$ versus $\kappa(b-a)/\h$ for the square barrier potential with varying dimensionless parameter $\omega_{\mathrm{A}}=a\sqrt{2mV_0}/\h$. The plot gives slope $\rho_{\mathrm{A}}$ as $\bigl\{(\omega_{\mathrm{A}},\rho_{\mathrm{A}})\bigr\}=\bigl\{(6,6.5062\pm 0.0002),(8,6.3027\pm 0.0002),(10,6.2086\pm 0.0002)\bigr\}$. Similarly, Figure \ref{twilightB} shows the plot for the modified square barrier potential with varying dimensionless parameter $\omega_{\mathrm{B}}=a\sqrt{2m(V_0-V_1)}/\h$. Here we set $V_0=3m,V_1=-1.5m,a=2\h/m$. The plot gives slope $\rho_{\mathrm{B}}$ as 
$\bigl\{(\omega_{\mathrm{B}},\rho_{\mathrm{B}})\bigr\}
=\bigl\{(6,8.4267\pm 0.0002),(8,8.2638\pm 0.0003),(10,8.1892\pm 0.0003)\bigr\}$. For P\"oschl-Teller potential $V(x)=U_0/\cosh^2[\alpha(x-b)]$ in 
Appendix~\ref{P\"oschl-Teller potential} we have,
\begin{equation}
		t_\circletophalfblack\approx\frac{4\gamma}{\Gamma_{0}}\frac{\pi\kappa}{\alpha\h}
\,,\quad
\label{twilight time: estimate 2}
\end{equation}
for a thick-barrier $\kappa/(\alpha\h)\gg1$ $\bigl(\kappa=\sqrt{2mU_0-k_0^2}$, $k_0$ is obtained from \eqref{k_0C}$\bigr)$ and large $b$ limit, $\gamma=1/2$. The plot of $\Gamma_{0}t_\circletophalfblack$ versus $\ln(\alpha/k_0)$ with varying dimensionless parameter $\omega_{\mathrm{C}}=b\sqrt{2mU_0}/\h$ is displayed in Figure~\ref{twilight2}. The plot gives slope $\rho_{\mathrm{C}}$ as $\bigl\{(\omega_{\mathrm{C}},\rho_{\mathrm{C}})\bigr\}=\bigl\{(40,1.670\pm0.007),(100,1.9607\pm0.0002),(200,1.9822\pm0.0001)\bigr\}$. Note that the slope in Figs.~\ref{twilightAB} and~\ref{twilight2} depends mostly on $v$, and only weakly on other dimensionless parameters.  In all 3 cases plots converge for large $\omega$, indicating the physical picture that for a barrier located far away from the origin, the product of oscillation period $T$ and decay rate $\Gamma_0$ match the tansmission coefficients $T_{\rm trans}$,
see the Appendices.

\section*{Dawn time}

The dawn time $t_\circlebottomhalfblack$ is the mnemonic term we propose for the point in time when
the exponential decay kicks in. To describe tunneling,
we pick an initial state that is localized in the left region.
We construct a superposition of eigenstates with support mainly in the left region. The initial wavefunction can be written as,
\begin{equation}
	\Psi(t=0,x)=\int_{0}^{\infty}\! \frac{\d p}{2\pi\h}\; c(p)\,\phi^{L}_{p}(x)
\,,
	\qquad
	\int_{0}^{\infty}\! \frac{\d p}{2\pi\h}\,|c(p)|^{2}=1
\,,\quad
\label{eq:init-continuum}
\end{equation}
\begin{equation}
	c(p)=\int_{0}^{\infty}\! \d x\; \phi^{L}_{p}(x)\,\Psi(0,x)
\,,\quad
\label{c(p)}
\end{equation}
with the momentum eigenvalue \(p\in(0,\infty)\). We then convolve this state with the kernel,
\begin{equation}
	\Psi(t,x)=\int_{0}^{\infty} \d y\, K(t,x;\,0,y)\,\Psi(0,y)
\,,\quad
\end{equation}
and using the late-time form of the 
kernel~\eqref{late time kernel} we obtain,
\begin{equation}
	\Psi(t,x)
	\approx  \int_{0}^{\infty}\! \frac{\d p}{2\pi\h}\; \!\int_{0}^{\infty}\!\d y\,
	c(p)\Big(
	G(E_0,x,y)\,(E_0 t)^{-\gamma}
	+ \sum_{n}F(E_n,x,y)\,\exp\!\Bigl\{-\tfrac{\Gamma_{n}}{2}\,t\Bigr\}
	\Big)\,\phi^{L}_{p}(y)
\,.\quad
\end{equation}
The power-law tail is obtained by using an asymptotic expansion valid at late time. The higher-order poles have a
larger imaginary component and will therefore decay quickly. The dawn time sets in when
the leading-order pole starts dominating the decay over the higher-order pole contributions, {\it i.e.} when
\begin{equation}
	\frac{\widetilde F(E_{0})\,\exp\!\{-\Gamma_{0}\, t\}}
	{\widetilde F(E_{n})\,\exp\!\{-\Gamma_{n}\, t\}}
	\gg 1 \qquad \left(\forall\, n\in\{1,2,3,\dots\}\right)
\,,\quad
\label{dawn}
\end{equation}
where we have introduced,
\[
\widetilde F(E_n)=\int_{\mathrm{L}}\!\mathrm{d}x\,\Biggl|\,
\int_{0}^{\infty}\!\int_{0}^{\infty}\!\mathrm{d}y c(p)\;F(E_n,x,y)\,\phi^{L}_{p}(y)
\Biggr|^{2}
\,,\quad
\]
where we have neglected the interference terms. Typically, some $\widetilde F(E_{n})$ in Eq.~\eqref{dawn} will dominate. Based on this consideration, we can extract the
following initial estimate for the dawn time,
\begin{equation}
	t_\circlebottomhalfblack
	= \max_{n}\left\{
	\frac{1}{\Gamma_{n}-\Gamma_{0}}
	\ln\!\left[
	\frac{\widetilde F(E_{n})}{\widetilde F(E_{0})}
	\right]\right\}
\,,\quad
\label{naive dawn time}
\end{equation}
where the initial-state dependence is captured by $\widetilde F(E_{n})$. Nevertheless, for some low energy initial states the higher modes $\Gamma_{n} \,(n\ge 1)$ cannot be excited, such that all terms in \eqref{naive dawn time} are negative.
In such cases, the dawn time is obtained by investigating when the $\Gamma_{0}$ exponential starts to dominate over the power-law tail, {\it i.e.} the earlier intersection point of~\eqref{twilight}, which is obtained from~\eqref{W0}. Hence, the final expression for the dawn time is,
\begin{equation}
	t_\circlebottomhalfblack
	=\max\left\{\max_{n}\left\{
	\frac{1}{\Gamma_{n}-\Gamma_{0}}
	\ln\!\left[
	\frac{\widetilde F(E_{n})}{\widetilde F(E_{0})}
	\right]\right\}, -\,\frac{2\gamma}{\Gamma_{0}}\,
	W_{0}\!\left(
	-\,\frac{\h}{E_{0}}
	\Bigg|\frac{G}{F}\Bigg|^{\!\frac{1}{\gamma}}\!
	\frac{\Gamma_{0}}{2\gamma}
	\right) \right\}
\;.\quad
\label{dawn time: estimate}
\end{equation}

To summarize, our analysis shows that the standard  expression for exponential decay of trapped states
(obtained {\it e.g.} by Euclidean methods) holds in the interval,
$t_\circlebottomhalfblack\leq t\leq t_\circletophalfblack$.

\section*{Square barrier example}

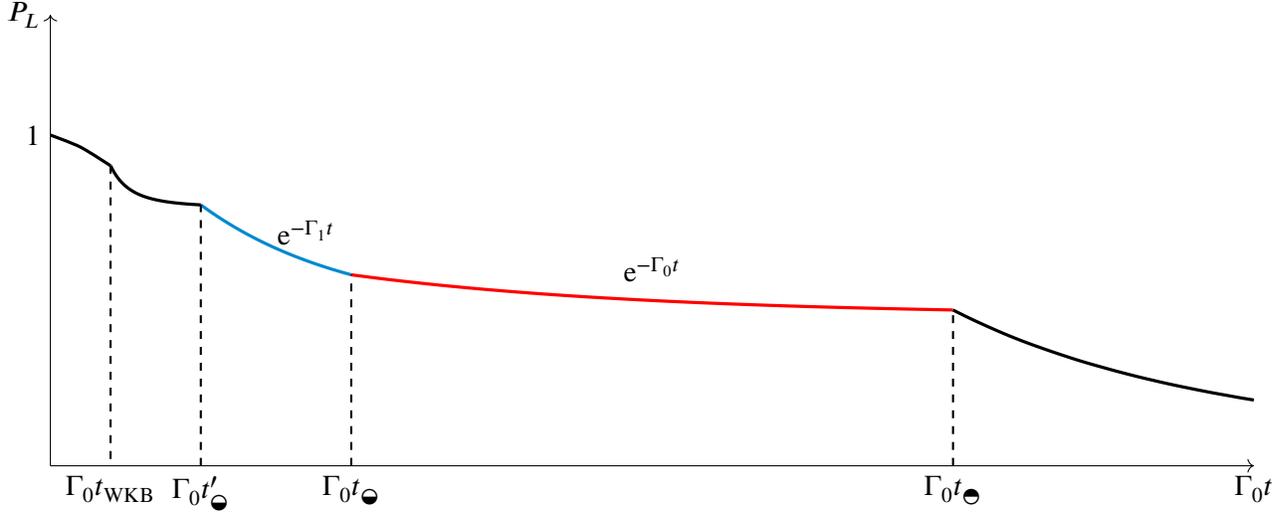
\begin{figure}
		\begin{tikzpicture}[scale=4]
		\draw[->] (0,-0.5)--(4,-0.5);
		\draw[->] (0,-0.5)--(0,1);
		\draw[very thick,domain =0.5:1,smooth,cyan!70!blue] plot(\x,{exp(-2*(\x))});
		\draw[very thick,domain =1:3,smooth,red] plot(\x,{0.3678*exp(-1*(\x))});
		\draw[very thick,domain =0.2:0.5,smooth] plot(\x,{0.00111*(\x)^(-3)+0.35893});
		\draw[very thick,domain =3:4,smooth] plot(\x,{14*(\x)^(-3)-0.5});
		\draw[very thick]
		plot[smooth]
		coordinates{
			(0,0.6)
			(0.1,0.56)
			(0.2,0.498)
		};
		\coordinate[label=below:$ \Gamma_0 t $](t) at (4,-0.5);
		\coordinate[label=left:$ P_L $](P_L) at (0,1);
		\coordinate[label=left:$ 1 $](1) at (0,0.6);
		\coordinate[label=below:$ \Gamma_{0}t_\circlebottomhalfblack' $](t0) at (0.5,-0.5);
		\coordinate[label=below:$ \Gamma_{0}t_\circlebottomhalfblack $](t1) at (1,-0.5);
		\coordinate[label=below:$ \Gamma_{0}t_\circletophalfblack $](t2) at (3,-0.5);
		\coordinate[label=below:$ \Gamma_{0}t_{\mathrm{WKB}} $](twkb) at (0.2,-0.5);
		\coordinate[label=above:$ \e^{-\Gamma_{1} t} $](Gamma1) at (0.85,0.2);
		\coordinate[label=above:$ \e^{-\Gamma_{0} t} $](Gamma1) at (2,0.08);
		\draw[dashed, thick] (0.5,-0.5)--(0.5,0.3678);
		\draw[dashed,thick] (1,-0.5)--(1,0.1353);
		\draw[dashed, thick] (3,-0.5)--(3,0.018);
		\draw[dashed, thick] (0.2,0.498)--(0.2,-0.5);
	\end{tikzpicture}
	\caption{Qualitative time evolution of the probability $P_L$ of finding the wave function in the left region $L$ for the initial wave function $\Psi_{0}=\sqrt{2/a}\sin(2\pi\mu x/a)$.
    The relevant times $t_{\rm WKB}$, $t'_\circlebottomhalfblack$, $t_\circlebottomhalfblack$ and $t_\circletophalfblack$ are explained in the text.}
	\label{Schematic plot}
\end{figure}

Let us now explore a particular example of the square barrier potential in Appendix~\ref{suqre barrier}. Take the initial state localized in the left well L to be the normalized state,
\[
\Psi_0(x)\;=\;\sqrt{\frac{2}{a}}\;\sin\!\Big(\frac{\mu\pi x}{a}\Big)
\,,\qquad \left( \mu\in\{1,2,3,\dots\}\right)
\,,\quad
\]
in which $a$ is the distance between the left edge of the barrier and the origin, see 
figure~\ref{fig.1}. For the tunneling case, the particle energy should be smaller than the barrier height, {\it i.e.} $\frac{\pi^2\h^2\mu^2}{2ma^2}< V_0$. The condition under which we can use late time expansion of the kernel to calculate the dawn time is $t\gg\frac{2ma^2}{\mu^2\pi^2\h}$. Expanding the initial state in the left–region basis \(\phi^{L}_{p}(x)=\dfrac{2}{N_p}\sin(px/\h)\) we obtain,
\begin{equation}
	c(p)=\int_{0}^{a}\! \d x\; \phi^{L}_{p}(x)\,\Psi_{0}(x)=\frac{2}{N_p}\sqrt{\frac{2}{a}}(-1)^{\mu+1}\;\frac{\frac{a}{\mu}\pi\,\sin(p a/\h)}{\pi^{2}-\frac{p^2 a^2}{\mu^2\h^2}}
\,,\quad
\end{equation}
from which $\widetilde F(E_n)$ can be calculated. For a thick-barrier we have, 
\begin{equation}
	\widetilde F(E_n)\approx\left(\frac{a}{2}+\frac{\kappa\h}{2mV_0}\right)\frac{\kappa^2}{2mV_0{(b-a)}^2}\frac{2}{a}\frac{\frac{a^2}{\mu^2}\pi^2}{\left[\pi^{2}-\frac{p_{n}^2 a^2}{\mu^2\h^2}\right]^2} \e^{-4\kappa(b-a)/\h}
\;,\qquad
\end{equation}
where $V_0$ is the height of the barrier, $\kappa (b-a)/\h=\sqrt{2mV_0-p_0^2}\times(b-a)/\h$ is the WKB exponent, $b-a$ is the width of the barrier, $p_0$ is obtained from \eqref{k_0AB}, $p_n\approx\frac{(n+1)\h\pi}{a+\h/\kappa}$. For $n\ge\mu$, the denominator $\left[\pi^{2}-\frac{p_{n}^2a^2}{\mu^2\h^2}\right]^2\sim \mathcal{O}(n^4)$ depresses $\widetilde F(E_n)$, therefore we only consider $n\le\mu-1$. Using $W_0(\epsilon)=\epsilon+\mathcal{O}(\epsilon^2)$ for $\epsilon\to0$, we obtain the dawn time,
\begin{equation}
	t_\circlebottomhalfblack=\max\left\{\max_{0\le n\le\mu-1}\left\{\frac{1}{\Gamma_{n}-\Gamma_{0}}\ln\!\left[\frac{ \left(\pi^{2}-\frac{p_{0}^2 a^2}{\mu^2\h^2}\right)^2}{\left(\pi^{2}-\frac{p_{n}^2 a^2}{\mu^2\h^2}\right)^2}\right]\right\}, m\h\left[\frac{b-a}{2\sqrt{2\pi}\kappa p_0\h}\right]^{\frac{2}{3}}  \right\}
\,.\quad
\end{equation}
Specifically, for $\mu=1$ we have,
\begin{equation}
	t_\circlebottomhalfblack=m\h\left[\frac{b-a}{2\sqrt{2\pi}\kappa p_0\h}\right]^{\frac{2}{3}}
\,.\quad
\end{equation}
Consider next $\mu=2$, in which case we can construct a complete picture for the probability $P_L$ of finding the wave function in the left region $L$. The dawn time is,
\begin{equation}
	t_\circlebottomhalfblack=\frac{1}{\Gamma_{1}-\Gamma_{0}}\ln\!\left[\frac{ \left(\pi^{2}-\frac{p_{0}^2 a^2}{4\h^2}\right)^2}{\left(\pi^{2}-\frac{p_{1}^2 a^2}{4\h^2}\right)^2}\right]
\,,\quad
\end{equation}
and the twilight time is,
\begin{equation}
	t_\circletophalfblack=\frac{6\kappa(b-a)}{\Gamma_{0}\h}
\,.\quad
\end{equation}
Figure~\ref{Schematic plot} is a schematic plot showing the relevant time scales in the decay of $P_L$. For example, $t_\circlebottomhalfblack' $, the time when $ \e^{-\Gamma_{1} t}$ decay kicks in, can be evaluated by investigating when $ \e^{-\Gamma_{1} t} $ decay starts  dominating over the power-law tail. To achieve that we only need to replace $\Gamma_{0}, F(E_0,x,y)$ by $\Gamma_{1}, F(E_1,x,y)$ in~\eqref{twilight} and choose the earlier intersection point, which is,
\begin{equation}
	t_\circlebottomhalfblack'=m\h\left[\frac{b-a}{2\sqrt{2\pi}\kappa p_1\h}\right]^{\frac{2}{3}}
\,.\quad 
\label{t'}
\end{equation}
It can be easily verified that $t_\circlebottomhalfblack'\gg t_{\mathrm{WKB}} \simeq\frac{ma^2}{2\pi^2\h}$ for sufficiently thick barrier $\kappa(b-a)/\h\gg(\kappa a/\h)^2$, which displays self-consistency of our 
picture.~\footnote{Eq.~\eqref{t'} does not hold for a general initial state, for which $t_\circlebottomhalfblack'$ signifies the onset of exponential decay of the state, and can be characterized by $\sum_n c_n\e^{-\Gamma_n t}$ for some  subset of $n$. At $t_\circlebottomhalfblack$,  $c_0\e^{-\Gamma_{0}t}$ begins to dominate.} 
\\ Our decay picture mirrors a universal script. After a short non-exponential transient, the signal enters exponential windows controlled by a few resonant poles, precisely the 'quasinormal' stage seen in black-hole ringdowns~\cite{Varghese:2012mr,Berti_2009,PhysRevLett.123.111102,PhysRevX.9.041060,Berti:2009kk} and the exponential plateau of cold-atom tunneling~\cite{Wilkinson1997,PhysRevLett.96.163601}. At very late times the pole dominance fades and a power-law tail takes over. Though the physics differs, the time dependence looks the same. The shared mechanism is spectral: early interference, mid-time quasinormal exponentials, late-time branch-cut tails. Our 'dawn-twilight' timings simply pick where each piece takes the lead.
\\

\section{Discussion}

Our analysis of the dawn time~(\ref{dawn time: estimate}) and twilight time~(\ref{twilight time: estimate})--(\ref{twilight time: estimate 2}) in one-dimensional tunneling ties together several complementary views of quantum decay, from textbook transmission coefficients to modern real-time saddle-point methods. For the square barrier, we showed that the rate extracted from the real-time kernel obeys $\Gamma T = T_{\rm trans}$ in the thick-barrier limit, with $T$ the classical oscillation period and $T_{\rm trans}$ the standard transmission probability. This makes the Gamow picture precise and clarifies how $T_{\rm trans}$ sets both the height and temporal extent of the exponential plateau before the universal power-law tail takes over.

In spectral language, the decay is controlled by the isolated complex poles embedded in a continuum. This viewpoint extends naturally to periodic (Bloch) systems, where isolated resonances broaden into bands and interwell tunneling is encoded in the Bloch bandwidth~\cite{Ambrozinski:2013cosine}, a picture characterizing many condensed matter systems. At a more formal level, the pole-plus-branch decomposition of the kernel provides a simple laboratory for the complex-saddle technology using Picard-Lefschetz theory~\cite{Witten:2010cs} and the real-time analysis of tunneling in the Rosen-Morse potential~\cite{Feldbrugge:2025complex}. Our approach sheds light on the Euclidean bounce and direct methods, and shows that Euclidean methods apply between $t_\circlebottomhalfblack$ and $t_\circletophalfblack$.~\footnote{Garbrecht and Wagner~\cite{Garbrecht:2025alb} have recently arrived at the same conclusion by using different techniques. However, these authors have not attempted to estimate the range of validity of the Euclidean approximation.} 

Finally, we surmise that suitably defined dawn and twilight times should exist more generally -- from periodic quantum-mechanical barriers to metastable vacua in quantum field theory and cosmology -- whenever decay is controlled by isolated complex poles embedded in a continuum. A detailed study is needed to make these statements quantitative, and thus more rigorous.

\appendix	
\section{Square barrier}\label{suqre barrier}

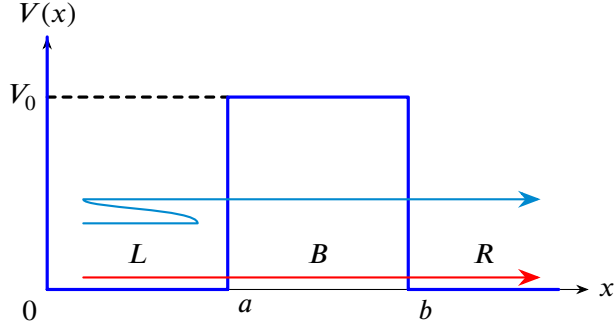
\begin{figure}[h]
	\centering
	\begin{tikzpicture}[>=Stealth, line cap=round, line join=round, x=0.8cm, y=0.8cm]
		\def\a{3.0}   
		\def\b{6.0}   
		\def\V{3.2}   
		\def\xmax{9} 
		\def\ymax{4.2}
		\draw[->] (0,0) -- (\xmax,0) node[right] {$x$};
		\draw[->] (0,0) -- (0,\ymax) node[above] {$V(x)$};
		\draw [dashed, very thick] (0,\V) -- (\a,\V);
		\draw[very thick, blue]
		(0,0) -- (0,\ymax)
		(0,0) -- (\a,0)
		(\a,0) -- (\a,\V) -- (\b,\V) -- (\b,0)
		(\b,0) -- (\xmax-0.5,0);
		\draw (\a,0) node[below right] {$a$};
		\draw (\b,0) node[below right] {$b$};
		\node[below left] at (0,0) {$0$};
		\node[left] at (0,\V) {$V_0$};
		\node at ({0.5*\a},0.6) {\small $L$};
		\node at ({0.5*(\a+\b)},0.6) {\small $B$};
		\node at ({0.5*(\b+\xmax-0.5)},0.6) {\small $R$};
		\draw[thick, cyan!70!blue, -{Stealth[length=3mm]}]
		(0.6,1.1)--(2.5,1.1) .. controls (2.5,1.35) and (0.6,1.35) .. (0.6,1.5)
		.. controls (0.6,1.5) and (\a,1.5) .. (\a,1.5)
		-- (\b,1.5) -- (\xmax-0.8,1.5);
		\draw[thick, red, -{Stealth[length=3mm]}]
		(0.6,0.20) -- (\a,0.20) -- (\b,0.20) -- (\xmax-0.8,0.20);
	\end{tikzpicture}
	\caption{Square potential barrier on the half line divided into three regions: $L$ (left), $B$ (barrier), and $R$ (the destination region to which a state initially localized in $L$ decays over time).}
	\label{fig.1}
\end{figure}

In this appendix, we discuss the decay rate for the simple square barrier on the half line shown in figure~\ref{fig.1}, and some generalizations. The potential is:
\begin{equation}
	V(x)=
	\begin{cases}
		+\infty\,, & x<0\,,\\
		0\,,       & 0\le x<a\,,\\
		V_0\,,     & a\le x<b\,,\\
		0\,,       & x\ge b\,.
	\end{cases}
\quad
\end{equation}
The energy eigenstates can be written as,
\begin{equation}\label{eigen-function-1}
	\phi_p(x) =
	\begin{cases}
		\phi_p^{L}(x) = \dfrac{2}{N_p}\sin(px/\h)\,, & 0 \le x < a\,, \\[8pt]
		\phi_p^{B}(x) = \dfrac{1}{N_p}\!\left[A_p \e^{\kappa(x-a)/\h} + B_p \e^{-\kappa(x-a)/\h}\right]\,, & a \le x < b\,, \\[8pt]
		\phi_p^{R}(x) = \dfrac{1}{N_p}\!\left[C_p \e^{\i p(x-b)/\h} + D_p \e^{-\i p(x-b)/\h}\right]\,, & x\ge b\,,\\[8pt]
	\end{cases}
\quad
\end{equation}
where $\kappa=\sqrt{2mV_0-p^2}$. The boundary condition $\phi(0)=0$ eliminates $\cos(px)$. Using continuity of $\psi$ and $\psi'$ we find,
\begin{align}
	A_p &= \sin(pa/\h) + \frac{p}{\kappa}\cos(pa/\h)\,,  \\[6pt]
	B_p &= \sin(pa/\h) - \frac{p}{\kappa}\cos(pa/\h)\,,  \\[6pt]
	C_p &= \tfrac{1}{2}\!\left(1 - \i\frac{\kappa}{p}\right) A_p \e^{W_p}
	+ \tfrac{1}{2}\!\left(1 + \i\frac{\kappa}{p}\right) B_p \e^{-W_p},  \\[6pt]
	D_p &= \tfrac{1}{2}\!\left(1 + \i\frac{\kappa}{p}\right) A_p \e^{W_p}
	+ \tfrac{1}{2}\!\left(1 - \i\frac{\kappa}{p}\right) B_p \e^{-W_p}
\,,\quad
\end{align}
where $W_p=\int_{a}^{b}\d x \,\kappa/\h$ is the WKB exponent. Here we consider a thick barrier, which requires $W_p\gg1$. The overall factor $N_p$ follows from the normalization condition,
\begin{equation}
	\int_{0}^{\infty} \d x \, \phi_{p}(x)\,\phi_{p'}^{*}(x)
	= \delta(p - p')
\,.\quad
\end{equation}
From Eq.~\eqref{eigen-function-1} we have,
\begin{align}
	\int_{0}^{\infty} \d x \, \phi_{p}(x)\,\phi_{p'}^{*}(x)
	&= \int_{b}^{\infty} \d x \, \phi_{p}^{R}(x)\,\phi_{p'}^{R*}(x)
	+ \int_{0}^{b} \d x \, \phi_{p}(x)\,\phi_{p'}^{*}(x) 
\notag\\[6pt]
	&= \pi \, \frac{C_{p}C_{p'}^{*} + D_{p}D_{p'}^{*}}
	{N_{p} N_{p'}^{*}} \, \delta(p-p')
\,.\quad
\end{align}
Noticing that $C_p=D_p^*$, we obtain
\begin{equation}
	{|N_p|}^2=2\pi C_pD_p
\;.\quad
\end{equation}

\subsection{Calculating the kernel}
\label{1.1}

To extract the decay rate, we first need the kernel $K(t,x;0,y)$ with $x,y\in L$. In path integral quantization the spectral representation reads,
\begin{align*}
	K(t,x;0,y)&=\int_{0}^{\infty} \frac{\d p}{2\pi\h} \,
	\phi_{p}^{*}(y)\,\phi_{p}(x)\,
	\e^{-\i\frac{p^{2}}{2m\h}t}\\
	&=\int_{0}^{\infty} \d E\left(\frac{1}{2\pi\h}\sqrt{\frac{m}{2E}}\right) \,
	\phi_{p}^{*}(y)\,\phi_{p}(x)\,
	\e^{-\i Et/\h}
\,,\quad
\end{align*}
and hence,
\begin{align}
	K(t,x;0,y)&=\int_{0}^{\infty} \d E\left(\frac{1}{2\pi\h}\sqrt{\frac{m}{2E}}\right)\frac{\sin\!\bigl(\sqrt{2mE}\,x/\h\bigr)\,
		\sin\!\bigl(\sqrt{2mE}\,y/\h\bigr)}
	{2\pi\,C(E)\,D(E)}\,
	\e^{-\i Et/\h}
\notag\\
	&=\int_{0}^{\infty} \d E\, F(t,x,y,E)
\,.\quad
\end{align}
We now analytically continue $|N(E)|^2=2\pi C(E)D(E)$ to a domain of complex energies and deform the contour to obtain,
\begin{equation}
	K(t,x;0,y)=f(t,x,y)+g(t,x,y)
\,,\quad
\end{equation}
where
\begin{equation}
	f(t,x,y)=\oint_{\mathcal{C}} \d E \,
	\left( \frac{1}{2\pi\h} \sqrt{\frac{m}{2E}} \right)
	\frac{ \sin\!\bigl(x\sqrt{2mE}/\h\bigr)\,
		\sin\!\bigl(y\sqrt{2mE}/\h\bigr) }
	{ 2\pi\,C(E)\,D(E) }\,
	\e^{-\i Et/\h}
\,,\quad
\end{equation}
and
\begin{align}
	g(t,x,y) &= \int_{0}^{-\i\infty} \d E \,
	\left( \frac{1}{2\pi\h} \sqrt{\frac{m}{2E}} \right)
	\frac{ \sin\!\bigl(x\sqrt{2mE}/\h\bigr)\,
		\sin\!\bigl(y\sqrt{2mE}/\h\bigr) }
	{ 2\pi\,C(E)\,D(E) }\,
	\e^{-\i Et/\h}\notag \\[8pt]
	&= -\,\i \int_{0}^{\infty} \d\mathcal{E} \,
	\left( \frac{1}{2\pi\h} \sqrt{\frac{m}{-\i\,2\mathcal{E}}} \right)
	\frac{ \sin\!\bigl(x\sqrt{-\i\,2m\mathcal{E}}/\h\bigr)\,
		\sin\!\bigl(y\sqrt{-\i\,2m\mathcal{E}}/\h\bigr) }
	{ 2\pi\,C(-\i\mathcal{E})\,D(-\i\mathcal{E}) }\,
	\e^{-\mathcal{E}t/\h}
\,.\quad
\end{align}
\begin{figure}[t]
	\centering
	\begin{tikzpicture}[scale=1.2, >=Stealth]
		\draw[->] (-2.2,0) -- (4.5,0) node[right] {$\mathrm{Re} E$};
		\draw[->] (0,-4) -- (0,2) node[above] {$\mathrm{Im} E$};
        \draw[black,very thick]
		plot[domain=-2.2:0, samples=180]
		(\x, {0.12*sin(900*\x)});
		\draw[black, very thick,->] (-2.2,0) -- (-2.32,0);
		\node[black,above] at (-1.35,0.18) {\scriptsize branch cut};
		\fill[black] (0,0) circle (2.2pt);
		\draw[red,thick,->] (0,-0.1) -- (4,-0.1) node[below right] {$\infty$};
		\draw[red,thick,->] (0,-4) -- (0,-0.1);
		\draw[red,thick,->] (4,-0.1) arc[start angle=0,end angle=-90,radius=4];
		\coordinate (pole0) at (1.5,-1.0);
		\coordinate (pole0*) at (1.5,1.0);
		\coordinate (pole1) at (2.5,-2);
		\coordinate (pole1*) at (2.5,2);
		\fill[red] (pole0) circle (2pt);
		\fill[red] (pole0*) circle (2pt);
		\fill[cyan!70!blue] (pole1) circle (2pt);
		\fill[cyan!70!blue] (pole1*) circle (2pt);
		\node[above left] at (pole0) {$E_0-\tfrac{\i}{2}\Gamma_0$};
		\node[above left] at (pole1) {$E_1-\tfrac{\i}{2}\Gamma_1$};
		\node at (3.0,-2.5) {$\mathcal{C}$};
		\node[right] at (0,-2.5) {$\mathcal{C}$};
		\node[below] at (3.0,-0.1) {$\mathcal{C}$};
	\end{tikzpicture}
	\caption{Contour $\mathcal{C}$ and poles of the kernel in the complex energy plane. The arc at infinity does not contribute because the integrand decays exponentially in this limit.}
\label{contour}
\end{figure}
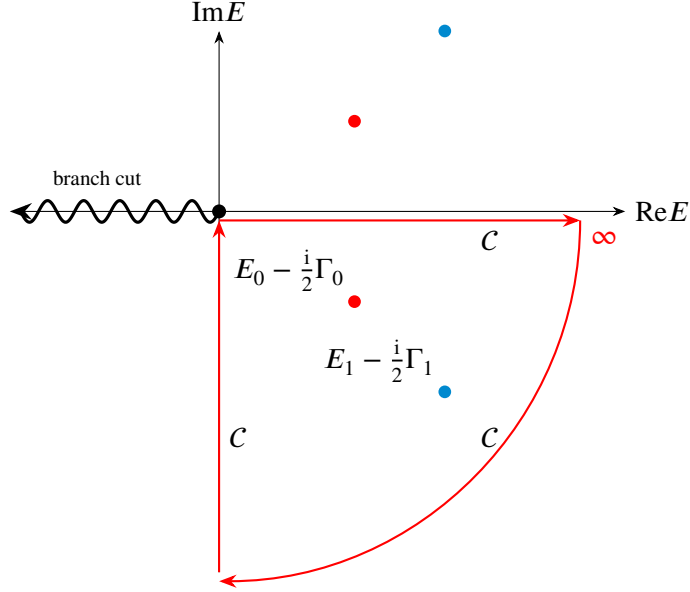
The contour $\mathcal{C}$ for the integral in $f(x,y,t)$ is shown in figure~\ref{contour}. The function $f(x,y,t)$ is evaluated by the residue theorem. The poles $E_n\!-\!\i b_n\,(n=0,1,\dots)$ are the (simple) zeros of $D(E)$, and hence 
\begin{align}
	f(t,x,y) 
	&= -2\pi \i \sum_{n} \operatorname{Res}\!\left(F,\, E_n \!-\! \i b_n \right) \notag\\[6pt]
	&\approx -\frac{\i}{2\pi\h} \sum_{n} 
	\sqrt{\frac{m}{2E_n}}\,
	\frac{ \sin\!\left( \sqrt{2mE_n} x/\h \right)\,
		\sin\!\left( \sqrt{2mE_n} y/\h \right) }
	{ C(E_n)\,D'(E_n) } \,\e^{-(\i E_nt+b_nt)/\h}\notag\\[8pt]
	&\approx -\frac{\i}{2\pi\h}\sqrt{\frac{m}{2E_0}}\,
	\frac{ \sin\!\left( \sqrt{2mE_0} x/\h \right)\,
		\sin\!\left( \sqrt{2mE_0} y/\h \right) }
	{ C(E_0)\,D'(E_0) } \,\e^{-(\i E_0t+b_0t)/\h}
\,,\quad
\end{align}
where the last approximation holds for $t \gg \hbar/b_0$ (late times). In this regime, we identify the decay rates associated with each pole as twice the imaginary part of the complex energy,
\begin{equation}
 \Gamma_n = 2 b_n/\h= -2\,\mathrm{Im}\{E\}/\h 
\,.\quad
\end{equation}

To evaluate $g(t,x,y)$, set $\mathcal{E} \to E_0 \alpha$~\footnote{Note that the choice of the rescaling is not unique. One could have also chosen $\mathcal{E} \to m \alpha$.} with dimensionless $\alpha$, such that
\begin{eqnarray}
	g(t,x,y) &\!=\!& -\i E_0 \left( \frac{1}{2\pi\h} \sqrt{\frac{m}{-\i\,2E_0}} \right)
\nonumber\\
	&&\hskip -1cm
\times\int_{0}^{\infty} \d\alpha \, \frac{1}{\sqrt{\alpha}}\,
	\frac{ \sin\!\bigl(\sqrt{-\i\,2mE_0}\,x\sqrt{\alpha}/\h\bigr)\,
		\sin\!\bigl(\sqrt{-\i\,2mE_0}\,y\sqrt{\alpha}/\h\bigr) }
	{ 2\pi\,C(-\i E_0\alpha)\,D(-\i E_0\alpha) }\,
	\e^{-\alpha(E_0 t)/\h}
\,.\qquad
\end{eqnarray}
We can expand $1(/C(-\i E_0\alpha)D(-\i E_0\alpha))$ near $0$
\[
\frac{1}{C(-\i E_0\alpha)\,D(-\i E_0\alpha)}
=\frac{1}{C(0)D(0)}\Bigl[1+(-\i E_0\alpha)\,s_E+\mathcal{O}(\alpha^2)\Bigr],\qquad
s_E:=C(0)D(0)\left(\frac{1}{C D}\right)'\!_{E}\Big|_{E=0}
\,.\quad
\]
\begin{equation*}
	s_E=\frac{\kappa_0\,|N(0)|\,Q-\kappa_0^{2}P^{2}}{V_0\,|N(0)|^{2}}
\,,\quad
\end{equation*}
where
\[
P=\frac{a}{\h}\cosh W_0+\frac{1}{\kappa_0}\sinh W_0,\qquad
Q=\frac{b}{\h}\sinh W_0+\frac{\kappa_0 a}{\h} d\cosh W_0
\,.\quad
\]
and $\kappa_0=\sqrt{2mV_0}$, $W_0=\kappa_0 (b-a)/\h$.

Set
\[
\beta =\sqrt{-\,\mathrm{i}\,2mE_0}/\h, \qquad \lambda = E_0 t/\h , \qquad u=\sqrt{\alpha}
\,.\quad
\]
and hence,
\begin{align*}
	\int_{0}^{\infty}\frac{\mathrm{d}\alpha}{\sqrt{\alpha}}\,
	\sin\!\bigl(\beta x\sqrt{\alpha}\bigr)\sin\!\bigl(\beta y\sqrt{\alpha}\bigr)\,\e^{-\lambda\alpha}
	&= \int_0^\infty \frac{2u\,\mathrm{d}u}{u}\,
	\sin(\beta x u)\sin(\beta y u)\,\e^{-\lambda u^2}\\
	&= 2\int_0^\infty \e^{-\lambda u^2}\,\sin(\beta x u)\sin(\beta y u)\,\mathrm{d}u\\
	&=\int_0^\infty \e^{-\lambda u^2}\Bigl[\cos\bigl(\beta(x-y)u\bigr)-\cos\bigl(\beta(x+y)u\bigr)\Bigr]\mathrm{d}u\\
	&=\frac{\sqrt{\pi}}{2\sqrt{\lambda}}\!
	\left[
	\exp\!\left(-\frac{(\beta(x-y))^2}{4\lambda}\right)
	-
	\exp\!\left(-\frac{(\beta(x+y))^2}{4\lambda}\right)
	\right]
\,.\quad
\end{align*}
Using also,
\[
\int_{0}^{\infty}\frac{\mathrm{d}\alpha}{\sqrt{\alpha}}\,
\alpha\,\sin\!\bigl(\beta x\sqrt{\alpha}\bigr)\sin\!\bigl(\beta y\sqrt{\alpha}\bigr)\,\e^{-\lambda\alpha}
=\frac{1}{2\lambda}\!
\left[
\exp\!\left(-\frac{(\beta(x-y))^2}{4\lambda}\right)
-
\exp\!\left(-\frac{(\beta(x+y))^2}{4\lambda}\right)
\right]
\,,\quad
\]
we obtain,
\begin{align*}
	g(t,x,y)&=-\mathrm{i}E_0\left(\frac{1}{2\pi\h}\sqrt{\frac{m}{-\mathrm{i}\,2E_0}}\right)\frac{1}{2\pi\,C(0)D(0)}\,
    \\
	&
	\times\Biggl\{
	\frac{1}{2}\sqrt{\frac{\pi\h}{E_0 t}}
	\Biggl[
	\e^{\mathrm{i}\frac{m(x-y)^2}{2t\h}}
	-
	\e^{\mathrm{i}\frac{m(x+y)^2}{2t\h}}
	\Biggr]\;
    \!+\!
	\frac{-\i E_0 s_E\h}{2 E_0 t}
	\Biggl[
	\e^{\mathrm{i}\frac{m(x-y)^2}{2t\h}}
	-
	\e^{\mathrm{i}\frac{m(x+y)^2}{2t\h}}
	\Biggr]
	\Biggr\}\\[4pt]
	&\approx-\mathrm{i}E_0\left(\frac{1}{2\pi\h}\sqrt{\frac{m}{-\mathrm{i}\,2E_0}}\right)\frac{1}{2\pi\,C(0)D(0)}\,
    \left(\frac{1}{2}\sqrt{\frac{\pi\h}{E_0 t}}-\frac{\i E_0 s_E\h}{2 E_0 t}\right)
	\\
	&\hskip 1.5cm 
    \times\Biggl[-\frac{2\i m}{t\h}xy+\Big(\frac{m}{t\h}\Big)^2 xy(x^2+y^2)+\mathcal{O}(t^{-3})\Biggr]
    \\[4pt]
	&= -(1+\i)(E_0 m)^{3/2} \,\frac{x y}{4\h^3\sqrt{\pi}\,|N(0)|^2}\,(E_0 t/\h)^{-3/2}+\mathcal{O}(t^{-2})
\,.\qquad
\end{align*}
In summary, the kernel scales as,
\begin{align}
	K(t,x;0,y) &\approx
	G(E_0,x,y)\,(E_0 t/\h)^{-3/2}
	+  \sum_{n} F(E_n,x,y)\,\exp\!\left\{-\tfrac{\Gamma_n}{2} t\right\} \notag \\[6pt]
	&\approx G(E_0,x,y)\,(E_0 t/\h)^{-3/2}
	+ F(E_0,x,y)\,\exp\!\left\{-\tfrac{\Gamma_0}{2} t\right\}
\,,\qquad
\end{align}
where the last line holds at late times. 
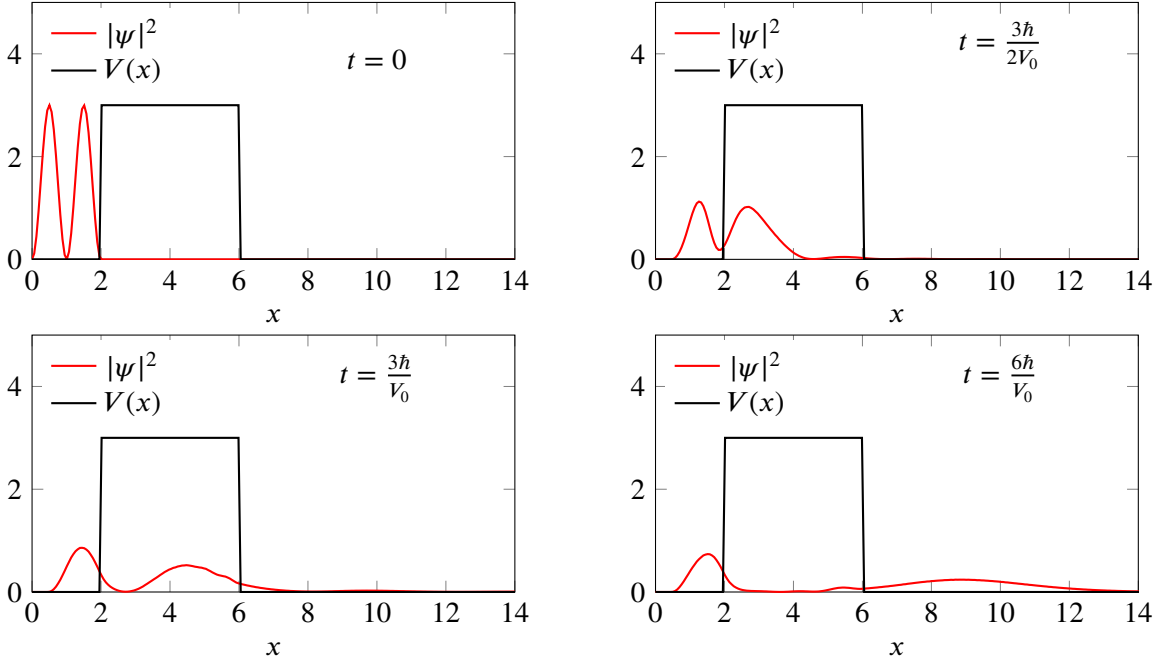
\begin{figure}[t!]
	\begin{subfigure}[t]{0.5\textwidth}
		\begin{tikzpicture}
			\begin{axis}[
				width=8cm, height=5cm,
				xmin=0, xmax=14, ymin=0, ymax=5,
				xlabel={$x$},
				legend style={draw=none, at={(0.02,0.98)}, anchor=north west},
				y tick label style={/pgf/number format/fixed}, scaled y ticks=false]
				\addplot+[no markers, thick, red]
				table[col sep=comma, header=true, x index=0, y index=1]{snap_t0.csv};
				\addlegendentry{$|\psi|^2$}
				\addplot+[no markers, thick, black]
				table[col sep=comma, header=true, x index=0, y index=2]{snap_t0.csv};
				\addlegendentry{$V(x)$}
				\node[above] at (10,350) {$t=0$};
			\end{axis}
		\end{tikzpicture}
	\end{subfigure}
	~
	\begin{subfigure}[t]{0.5\textwidth}
		\begin{tikzpicture}
			\begin{axis}[
				width=8cm, height=5cm,
				xmin=0, xmax=14, ymin=0, ymax=5,
				xlabel={$x$}, 
				legend style={draw=none, at={(0.02,0.98)}, anchor=north west},
				y tick label style={/pgf/number format/fixed}, scaled y ticks=false]
				\addplot+[no markers, thick, red]
				table[col sep=comma, header=true, x index=0, y index=1]{snap_t0p500.csv};
				\addlegendentry{$|\psi|^2$}
				\addplot+[no markers, thick, black]
				table[col sep=comma, header=true, x index=0, y index=2]{snap_t0.csv};
				\addlegendentry{$V(x)$}
				\node[above] at (10,350) {$t=\frac{3\h}{2V_0}$};
			\end{axis}
		\end{tikzpicture}
	\end{subfigure}
	\begin{subfigure}[t]{0.5\textwidth}
		\begin{tikzpicture}
			\begin{axis}[
				width=8cm, height=5cm,
				xmin=0, xmax=14, ymin=0, ymax=5,
				xlabel={$x$}, 
				legend style={draw=none, at={(0.02,0.98)}, anchor=north west},
				y tick label style={/pgf/number format/fixed}, scaled y ticks=false]
				\addplot+[no markers, thick, red]
				table[col sep=comma, header=true, x index=0, y index=1]{snap_t1.csv};
				\addlegendentry{$|\psi|^2$}
				\addplot+[no markers, thick, black]
				table[col sep=comma, header=true, x index=0, y index=2]{snap_t1.csv};
				\addlegendentry{$V(x)$}
				\node[above] at (10,350) {$t=\frac{3\h}{V_0}$};
			\end{axis}
		\end{tikzpicture}
	\end{subfigure}
	~
	\begin{subfigure}[t]{0.5\textwidth}
		\begin{tikzpicture}
			\begin{axis}[
				width=8cm, height=5cm,
				xmin=0, xmax=14, ymin=0, ymax=5,
				xlabel={$x$}, 
				legend style={draw=none, at={(0.02,0.98)}, anchor=north west},
				y tick label style={/pgf/number format/fixed}, scaled y ticks=false]
				\addplot+[no markers, thick, red]
				table[col sep=comma, header=true, x index=0, y index=1]{snap_t2.csv};
				\addlegendentry{$|\psi|^2$}
				\addplot+[no markers, thick, black]
				table[col sep=comma, header=true, x index=0, y index=2]{snap_t2.csv};
				\addlegendentry{$V(x)$}
				\node[above] at (10,350) {$t=\frac{6\h}{V_0}$};
			\end{axis}
		\end{tikzpicture}
	\end{subfigure}
	\caption{The numerical evolution of a particle initially localized in the left region. The unit of $x$ is $a/2$.}
\label{numerical decay}
\end{figure} 

\subsection{Explicit calculation of the rate}
\label{1.2}

To obtain an explicit decay rate, we need the roots of $D(E)$, which lie exponentially close to the roots of $A(E)$. For convenience, we use $p$ instead of $E$. Setting $A_p=0$ implies 
\begin{align}
	\sin(p_R a/\h) &= -\frac{p_R}{\sqrt{p_R^2 + \kappa_R^2}}\,, 
	& \quad
	\cos(p_R a/\h) &= \frac{\kappa_R}{\sqrt{p_R^2 + \kappa_R^2}}
\;,\quad
\label{k_0AB}
\end{align}
where $\kappa_R \equiv \sqrt{2mV_0 - p_R^2}$ and $p_R$ is the real resonance momentum. Here we can obtain an approximate solution for $p_{R,n}$,
\begin{equation}
	p_{R,n}\approx\frac{(n+1)\h\pi}{a+\h/\kappa_{R,0}}
\;,\quad
\end{equation}
We then find the complex zeros of $D_p$ by expanding perturbatively in,
\begin{equation}
	\delta \equiv \e^{-W_p} = \e^{-\kappa_R (b-a)/\h}.
\end{equation}
Writing $p = p_{R} + \delta^{2} p_{C} + \mathcal{O}(\delta^{4})$ and expanding $D_p$ to order $\delta$, the condition $D_p=0$ yields,
\begin{equation}
	p_{C} = \frac{2 p_{R} \kappa_{R}^{2}}
	{(p_{R} + \i \kappa_{R})^{2} (1 + a \kappa_{R}/\h)}
\;,\quad
\end{equation}
and thus,
\begin{equation}
\label{gamma}
	\Gamma = -\frac{2}{\h}\,\mathrm{Im}\!\left(\frac{p^{2}}{2m}\right)
	= \frac{8 p_{R}^{3} \kappa_{R}^{3}}
	{ m\h (1 + a \kappa_{R}/\h) (p_{R}^{2} + \kappa_{R}^{2})^{2} }
	\, \e^{-2\kappa_R (b-a)/\h} 
\,.\quad
\end{equation}
To estimate the twilight time for this potential we also need,  
\begin{align}
	|F|
	&= \frac{1}{2\pi\h}\sqrt{\frac{m}{2E_0}}\;
	\frac{|\sin(p_0 x/\h)\,\sin(p_0 y/\h)|}{\,|C(E_0)\,D'(E_0)|\,}\,,\\[4pt]
	|G|
	&= \frac{(mE_0)^{3/2}}{2\sqrt{2\pi}\h^3\,|N(0)|^2}\;|x\,y|\,,
	\qquad
	|N(0)|^2=2\pi\,C(0)\,D(0)
\;,\quad
\end{align}
and hence,
\begin{equation}
	\left|\frac{G}{F}\right|
	=
	\frac{mE_0^2}{2\sqrt{\pi}}\;
	\frac{|C(E_0)\,D'(E_0)|}{\,C(0)\,D(0)}\frac{|x\,y|,}{|\sin(p_0 x)\,\sin(p_0 y)|}
\;.\quad
\end{equation}
For small $x,y$ we then have
\begin{equation}
	\left|\frac{G}{F}\right|
	\approx
	\frac{mE_0^2}{2\sqrt{\pi}p_0^2}\;
	\frac{|C(E_0)\,D'(E_0)|}{\,C(0)\,D(0)}
\,.\quad
\end{equation}
Together with,
\begin{equation}
	\Gamma_0 \;=\;
	\frac{8\,p_0^{3}\,\kappa^{3}}
	{\h m\,(1+a\kappa/\h)\,\bigl(p_0^{2}+\kappa^{2}\bigr)^{2}}\;
	e^{-2\kappa (b-a)/\h}
\;,\quad
\end{equation}
we finally obtain,
\begin{equation}
	t_\circletophalfblack^{(A)}=\;
	-\,\frac{3}{\Gamma_0}\;
	W_{-1}\!\left(
	-\,\frac{\Gamma_0\h}{3E_0}\;
	\left[
	\frac{p_0^2}{8m\sqrt{\pi}}\;
	\frac{|C(E_0)\,D'(E_0)|}{\,C(0)\,D(0)}
	\right]^{\!\!2/3}
	\right)
\;.\quad
	\label{eq:t2-SB}
\end{equation}
For a thick-barrier case,  using $W_{-1}(-\epsilon)=\ln(\epsilon)+\mathcal{O}(\ln|\ln\epsilon|)$ for $\epsilon\to0$, this can be approximated by,
\begin{equation}
	t_\circletophalfblack\approx\frac{6\kappa(b-a)}{\Gamma_0\h}
\;,\quad
\end{equation}
which is Eq.~(\ref{twilight time: estimate}) with $\gamma=3/2$.

\subsection{Physical interpretation}\label{1.3}

A semi-classical picture helps to relate the poles to the decay rate. Poles of different order correspond to different modes of oscillation in the well that subsequently tunnel through the barrier. For example, the red and blue poles in Figure~\ref{contour} correspond to the colored schematic trajectories in figure~\ref{fig.1}, and snapshots of the actual decay are shown in figure~\ref{numerical decay}.

We can verify this picture in an extreme case: as $a \to +\infty$, the square barrier on the half line becomes a square barrier on the whole line, a standard example in quantum mechanics. From Eq.~\eqref{gamma} we obtain
\begin{equation}
	\lim_{a\to+\infty}\Gamma=\frac{8 p_{R}^{3} \kappa_{R}^{2}}
	{ m a (p_{R}^{2} + \kappa_{R}^{2})^{2} }
	\, \e^{-2W}.
\end{equation}
For such a case, only the lowest order of poles (the red trajectory) matters. Multiplying $\Gamma$ by one oscillation period $T=\dfrac{2ma}{p_R}$, a transmission coefficient $T_{\rm trans}$ emerges:
\begin{equation}
	T_{\rm trans}=T\,\Gamma=\frac{16 p_{R}^{2} \kappa_{R}^{2}}
	{ (p_{R}^{2} + \kappa_{R}^{2})^{2} }
	\, \e^{-2W}
\,,\quad
\end{equation}
which matches the thick-barrier limit of the transmission coefficient for a square barrier\cite{LANDAU197750},
\begin{align}
	\lim_{\kappa (b-a)\gg1}T_{\rm trans}&=\lim_{\kappa (b-a)\gg1}\frac{1}{1 + \dfrac{(p^{2} + \kappa^{2})^{2}}{4p^{2}\kappa^{2}} \sinh^{2}(\kappa (b-a)/\h)}\notag\\
	&=\frac{16 p^{2} \kappa^{2}}
	{ (p^{2} + \kappa^{2})^{2} }
	\, \e^{-2\kappa (b-a)/\h}
\;.\quad
\end{align}

\section{Modified square barrier}
\label{Modified square barrier}

\begin{figure}[h]
	\centering
	\begin{tikzpicture}[>=Stealth, line cap=round, line join=round, x=0.8cm, y=0.8cm]
		\def\a{3.0}
		\def\b{6.0}
		\def\V{3.2}
		\def\W{-1}
		\def\xmax{9}
		\def\ymax{4.2}
		\draw[->] (0,0) -- (\xmax,0) node[right] {$x$};
		\draw[->] (0,-1.2) -- (0,\ymax) node[above] {$V(x)$};
		\draw[very thick, blue]
		(0,0) -- (0,\V)
		(0,0) -- (\a,0)
		(\a,0) -- (\a,\V) -- (\b,\V) -- (\b,\W)
		(\b,\W) -- (\xmax-0.5,\W);
        \draw[very thick, dashed]
        (0,\V) -- (\a,\V)
        (0,\W) -- (\b,\W);
		\draw (\a,0) node[below right] {$a$};
		\draw (\b,0) node[below right] {$b$};
		\node[below left] at (0,0) {$0$};
		\node[left] at (0,\V) {$V_0$};
		\node[left] at (0,\W) {$V_1$};
		\node at ({0.5*\a},0.6) {\small $L$};
		\node at ({0.5*(\a+\b)},0.6) {\small $B$};
		\node at ({0.5*(\b+\xmax-0.5)},\W+0.6) {\small $R$};
	\end{tikzpicture}
	\caption{Modified square barrier.}
	\label{fig.3}
\end{figure}
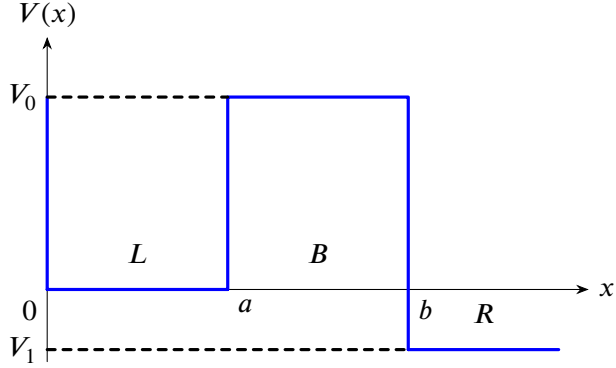

Let us consider a more realistic model of false vacuum decay, the modified square barrier, which is analogous to the MIT Bag Model~\cite{Chodos:1974je},
\begin{equation}
	V(x)=
	\begin{cases}
		+\infty, & x<0\,,\\
		0\,,       & 0\le x<a\,,\\
		V_0\,,     & a\le x<b\,,\\
		V_1\,,    & x\ge b\,,
	\end{cases}
\quad
\end{equation}
shown in Figure~\ref{fig.3}. Here we consider $V_0>0, V_1<0$. The eigenstates are,
\begin{equation}\label{eigen-function-2}
	\phi_p(x) =
	\begin{cases}
		\phi_p^{L}(x) = \dfrac{2}{N_p}\sin(px/\h)\,, & 0 \le x < a\,, \\[8pt]
		\phi_p^{B}(x) = \dfrac{1}{N_p}\!\left[A_p \e^{\kappa(x-a)/\h} + B_p \e^{-\kappa(x-a/\h}\right]\,, & a \le x < b\,, \\[8pt]
		\phi_p^{R}(x) = \dfrac{1}{N_p}\!\left[C_p \e^{\i q(x-b)/\h} + D_p \e^{-\i q(x-b)/\h}\right]\,, & x\ge b\,,\\[8pt]
	\end{cases}
\quad
\end{equation}
where $\kappa=\sqrt{2mV_0-p^2}$, $q=\sqrt{-2mV_1+p^2}$, and
\begin{align}
	A_p &= \sin(pa/\h) + \frac{p}{\kappa}\cos(pa/\h)\,,  \\[6pt]
	B_p &= \sin(pa/\h) - \frac{p}{\kappa}\cos(pa/\h)\,,  \\[6pt]
	C_p &= \tfrac{1}{2}\!\left(1 - \i\frac{\kappa}{q}\right) A_p \e^{W_p}
	+ \tfrac{1}{2}\!\left(1 + \i\frac{\kappa}{q}\right) B_p \e^{-W_p}\,,  \\[6pt]
	D_p &= \tfrac{1}{2}\!\left(1 + \i\frac{\kappa}{q}\right) A_p \e^{W_p}
	+ \tfrac{1}{2}\!\left(1 - \i\frac{\kappa}{q}\right) B_p \e^{-W_p}
\,.\quad
\end{align}
Using the normalization condition we have,
\begin{equation}
	\int_{0}^{\infty} \d x \, \phi_{p}(x)\,\phi_{p'}^{*}(x)
	= \pi \, \frac{C_{p}C_{p'}^{*} + D_{p}D_{p'}^{*}}
	{N_{p} N_{p'}^{*}} \, \delta(p-p')= \delta(p - p')
\,.\quad
\end{equation}
For $\delta(q-q')=\dfrac{q}{p}\delta(p-p')$ we obtain, 
\begin{equation}
	{|N_p|}^2=2\pi C_pD_p\sqrt{\frac{-V_1+E}{E}}
\,.\quad 
\end{equation}

Proceeding as before,
\begin{equation}
	K(t,x;0,y)=f(t,x,y)+g(t,x,y)
\,,\quad
\end{equation}
where $f(t,x,y)$ remains the same as in subsection~\ref{1.1}, while $g(t,x,y)$ differs,
\begin{align}
	g(t,x,y) &= \int_{0}^{-\i\infty} \d E \,
	\left( \frac{1}{2\pi/\h} \sqrt{\frac{m}{2(-V_1+E)}} \right)
	\frac{ \sin\!\bigl(x\sqrt{2mE}/\h\bigr)\,
		\sin\!\bigl(y\sqrt{2mE}/\h\bigr) }
	{ 2\pi\,C(E)\,D(E) }\,
	\e^{-\i Et/\h}\notag \\[8pt]
	&\hskip -0.cm
    = -\,\i \int_{0}^{\infty} \d\mathcal{E} \,
	\left( \frac{1}{2\pi/\h} \sqrt{\frac{m}{-2V_1-\i\,2\mathcal{E}}} \right)
	\frac{ \sin\!\bigl(x\sqrt{-\i\,2m\mathcal{E}}/\h\bigr)\,
		\sin\!\bigl(y\sqrt{-\i\,2m\mathcal{E}}/\h\bigr) }
	{ 2\pi\,C(-\i\mathcal{E})\,D(-\i\mathcal{E}) }\,
	\e^{-\mathcal{E}t/\h}
\,.\quad
\end{align}
Introducing a rescaling $\mathcal{E} \to E_0 \alpha$, one obtains,
\begin{align}
	g(t,x,y) &= -\i E_0 \left( \frac{1}{2\pi/\h} \sqrt{\frac{m}{-2V_1-\i\,2E_0\alpha}} \right)
    \\
    &
	\times\int_{0}^{\infty} \d\alpha \,
	\frac{ \sin\!\bigl(\sqrt{-\i\,2mE_0}\,x\sqrt{\alpha}/\h\bigr)\,
		\sin\!\bigl(\sqrt{-\i\,2mE_0}\,y\sqrt{\alpha}/\h\bigr) }
	{ 2\pi\,C(-\i E_0\alpha)\,D(-\i E_0\alpha) }\,
	\e^{-\alpha(E_0 t)/\h}\notag
\,.\quad
\end{align}
We now expand $1/(C(-\i E_0\alpha)D(-\i E_0\alpha))$ to linear order in $\alpha$ {\it via},
\[
\frac{1}{C(-\i E_0\alpha)D(-\i E_0\alpha)}
=\frac{1}{C(0)D(0)}\Bigl[1+(-\i E_0\alpha)\,s_E+\mathcal{O}(\alpha^2)\Bigr],
\qquad
s_E=\left(\frac{1}{CD}\right)'\!_{E}\Big|_{E=0},
\]
\begin{equation*}
	s_E=\frac{\kappa_0\,\sqrt{C(0)D(0)}\,Q-\kappa_0^{2}P^{2}}{V_0\,(C(0)D(0))^{2}},
\end{equation*}
where
\[
P=\frac{a}{\h}\cosh W_0+\frac{1}{\kappa_0}\sinh W_0,\qquad
Q=\frac{b}{\h}\sinh W_0+\kappa_0 a d/\h\cosh W_0.
\]
and $\kappa_0=\sqrt{2mV_0}$, $W_0=\kappa_0 (b-a)/\h$. 
Set $ \lambda= E_0 t/\h,\quad
\beta= -2V_1,\quad \gamma= \mathrm{i}\,2E_0,\quad u=\sqrt{\alpha}
$, and then we have,
\begin{align*}
	g(t,x,y)&=
	-\mathrm{i}\,E_0\frac{\sqrt{m}}{4\pi^2\h C(0)D(0)}
    \\
    &\hskip 0.2cm
    \times\int_{0}^{\infty} 2u\Bigl[1+(-\i E_0 u^2)\,s_E+\mathcal{O}(\alpha^2)\Bigr]\,
    \frac{\sin(\sqrt{-\,\mathrm{i}\,2mE_0} x u)\sin(\sqrt{-\,\mathrm{i}\,2mE_0} y u)}{\sqrt{\beta-\gamma u^2}}\,\mathrm{e}^{-\lambda u^2}\,\mathrm{d}u\\
	&\hskip -0.cm
    =
	-\mathrm{i}\,E_0\frac{\sqrt{m}}{4\pi^2\h C(0)D(0)}
	\int_{0}^{\infty}
	\frac{u(1-\i E_0 u^2\,s_E)\,\mathrm{e}^{-\lambda u^2}}{\sqrt{\beta-\gamma u^2}}\,
	\Big[\cos(b_- u)-\cos(b_+ u)\Big]\,\mathrm{d}u
\,,\quad
\end{align*}
where $b_\pm=\sqrt{-\,\mathrm{i}\,2mE_0}(x\mp y)$.

For $\mathrm{Re}[\beta]>0$,
\[
\frac{1}{\sqrt{\beta-\gamma u^2}}
=\frac{1}{\sqrt{\pi}}\int_{0}^{\infty} s^{-1/2}\,\mathrm{e}^{-\beta s}\,\mathrm{e}^{\gamma s\,u^2}\,\mathrm{d}s
\,.\quad
\]
we obtain,
\begin{align*}
	g(t,x,y)&=
	-\i E_0\,\frac{\sqrt{m}}{(2\pi)^2\,C(0)D(0)\,\sqrt{\pi}}
	\int_{0}^{\infty}\! \d s\; s^{-1/2}\,\e^{-\beta s}\,
	\Big[F(b_-;p)-F(b_+;p)\\[-2pt]
	&\hspace{6.7cm}\quad+\,(-\i E_0 s_E)\,\big(H(b_-;p)-H(b_+;p)\big)\Big]
\,,\quad
\end{align*}
where $p=\lambda-\i\,2E_0 s$ and
\begin{align*}
	F(b;p)&=\int_0^\infty u\,\e^{-p u^2}\cos(bu)\,\d u
	=\frac{1}{2p}-\frac{\sqrt{\pi}\,b}{4\,p^{3/2}}\,
	\exp\!\Big(-\frac{b^2}{4p}\Big)\,
	\mathrm{erfi}\!\Big(\frac{b}{2\sqrt{p}}\Big)\\
	&=\frac{1}{2p}-\frac{b^2}{4\,p^{2}}+\frac{b^4}{24\,p^{3}}-\frac{b^6}{240\,p^{4}}+\cdots,\\[4pt]
	H(b;p)&=\int_0^\infty u^{3}\,\e^{-p u^2}\cos(bu)\,\d u
	=-\frac{\partial}{\partial p}F(b;p)
	=\frac{1}{2p^{2}}-\frac{b^2}{2p^{3}}+\frac{b^4}{6p^{4}}-\frac{b^6}{24p^{5}}+\cdots
\,,\quad
\end{align*}
and hence,
\begin{equation*}
	F(b_-;p)-F(b_+;p)=
	-\frac{b_-^2-b_+^2}{4p^2}
	+\frac{b_-^4-b_+^4}{24p^3}
	-\frac{b_-^6-b_+^6}{240p^4}
	+\cdots,
\end{equation*}
and
\begin{equation*}
	H(b_-;p)-H(b_+;p)=
	-\frac{b_-^2-b_+^2}{2p^3}
	+\frac{b_-^4-b_+^4}{6p^4}
	-\frac{b_-^6-b_+^6}{24p^5}
	+\cdots
\,.\quad
\end{equation*}
For $p=\lambda-\i 2E_0 s$ and integer $n\ge1$,
\begin{equation*}
	p^{-n}
	=\lambda^{-n}\Big(1-\frac{\i 2E_0 s}{\lambda}\Big)^{-n}
	=\lambda^{-n}\sum_{j=0}^{\infty}\binom{n+j-1}{j}
	\Big(\frac{\i 2E_0 s}{\lambda}\Big)^{\!j}
\,,\quad
\end{equation*}
and thus
\[
\int_{0}^{\infty}\d s\, s^{j-\frac12}\e^{-\beta s}
=\frac{\Gamma\!\big(j+\tfrac12\big)}{\beta^{\,j+\frac12}},\qquad \beta=-2V_1
\,.\quad
\]
Hence, up to $\mathcal{O}(t^{-3})$,
\begin{align*}
	g(t,x,y)
	&=-\sqrt{\frac{-V_1+E_0}{-V_1}} \,\frac{x y( E_0m)^{3/2}}{\sqrt{2}\pi\h^3\,|N(0)|^2}\,(E_0 t/\h)^{-2}\\
	&\quad
	+\;\i\sqrt{\frac{-V_1+E_0}{-V_1}}\frac{E_0}{V_1} \,\frac{x y\big(3-2mV_1(x^2+y^2)/\h^2\big)( E_0m)^{3/2}}{3\sqrt{2}\pi\h^3\,|N(0)|^2}\,(E_0 t/\h)^{-3}\\
	&\quad
	+\;\i\sqrt{\frac{-V_1+E_0}{-V_1}}\frac{E_0}{V_1}\,\frac{(E_0 m)^{3/2}}{\sqrt{2}\pi}\,
	\frac{s_E}{|N(0)|^2}\;x y\,(E_0 t/\h)^{-3}
	+\;O(t^{-4})
\,.\quad
\end{align*}
Therefore the kernel scales as:
\begin{align}
K(t,x;0,y) &\approx
	G(E_0,x,y)\,(E_0 t/\h)^{-2}
	+  \sum_{n} F(E_n,x,y)\,\exp\!\left\{-\tfrac{\Gamma_n}{2} t\right\} \notag \\[6pt]
	&\approx G(E_0,x,y)\,(E_0 t/\h)^{-2}
	+ F(E_0,x,y)\,\exp\!\left\{-\tfrac{\Gamma_0}{2} t\right\}
\,.\quad
\end{align}
The explicit decay rate follows from the same method as in subsection~\ref{1.2},
\begin{equation}\label{gamma2}
	\Gamma= \frac{8 p_{R}^{2} q_R \kappa_{R}^{3}}
	{ \h m (1 + a \kappa_{R}/\h) (p_{R}^{2} + \kappa_{R}^{2})(q_{R}^{2} + \kappa_{R}^{2}) }
	\, \e^{-2W}
\,.\quad
\end{equation}
In the large-$a$ limit,
\begin{equation}
	\lim_{a\to+\infty}\Gamma=\frac{8 p_{R}^{2} q_R \kappa_{R}^{2}}
	{ m a (p_{R}^{2} + \kappa_{R}^{2})(q_{R}^{2} + \kappa_{R}^{2}) }
	\, \e^{-2W}
\,,\quad
\end{equation}
and therefore,
\begin{equation}
	T_{\rm trans}=T\,\Gamma=\frac{16 p_{R} q_R \kappa_{R}^{2}}
	{ (p_{R}^{2} + \kappa_{R}^{2})(q_{R}^{2} + \kappa_{R}^{2}) }
	\, \e^{-2W}
\,,\quad
\end{equation}
which matches the thick-barrier limit with the corresponding transmission coefficient,
\begin{align}
	\lim_{\kappa (b-a)\gg1}D&=\lim_{\kappa (b-a)\gg1}\frac{4 p q \kappa^{2}}
	{(pq - \kappa^{2})^{2} \sinh^{2}(\kappa (b-a)/\h) 
		+ \kappa^{2}(p+q)^{2} \cosh^{2}(\kappa (b-a)/\h)}\notag\\
	&=\frac{16 p q \kappa^{2}}
	{ (p^{2} + \kappa^{2})(q^{2} + \kappa^{2}) }
	\, \e^{-2\kappa (b-a)/\h}
\,.\quad
\end{align}
Using similar method as in Appendix~\ref{suqre barrier}, we can estimate the twilight time, 
\[ |G|=\sqrt{\frac{E_0}{2V_1}}\frac{(E_0 m)^{3/2} xy}{2\pi^2 C(0)D(0)\h^3},\qquad |F|
= \frac{1}{2\pi\h}\sqrt{\frac{m}{2E_0}}\;
\frac{|\sin(p_0 x/\h)\,\sin(p_0 y/\h)|}{\,|C(E_0)\,D'(E_0)|\,} \]
\begin{equation}
	\left|\frac{G}{F}\right|
	\;\approx\;
	\frac{p_0^3}{4\sqrt{2V_1}\pi m^{3/2}}\;
	\frac{|C(E_0)\,D'(E_0)|}{\,C(0)\,D(0)}
\,,\quad
\end{equation}
and
\begin{equation}
	\Gamma_0 \;=\;
	\frac{8\,p_0^{2}\,q\,\kappa^{3}}
	{m\,(1+a\kappa)\,\bigl(p_0^{2}+\kappa^{2}\bigr)\,\bigl(q^{2}+\kappa^{2}\bigr)}\;
	e^{-2\kappa (b-a)/\h}
\,,\quad
\end{equation}
and hence
\begin{equation}
	t_\circletophalfblack^{(B)}\;=\;
	-\,\frac{4}{\Gamma_0}\;
	W_{-1}\!\left(
	-\,\frac{\Gamma_0\h}{4E_0}\;
	\left[
	\frac{p_0^3}{4\sqrt{2V_1}\pi m^{3/2}}\;
	\frac{|C(E_0)\,D'(E_0)|}{\,C(0)\,D(0)}
	\right]^{\!\!1/2}
	\right)
\,.\quad
	\label{eq:t2-mSB}
\end{equation}
For a thick-barrier case, using $W_{-1}(-\epsilon)=\ln(\epsilon)+\mathcal{O}(\ln|\ln\epsilon|)$ for $\epsilon\to0$, we have
\begin{equation}
	t_\circletophalfblack\approx\frac{8\kappa_0 (b-a)}{\Gamma_0\h}
\,,\quad
\end{equation}
which agrees with Eq.~(\ref{twilight time: estimate}) with $\gamma=2$.

\section{P\"oschl-Teller potential}
\label{P\"oschl-Teller potential}

\begin{figure}[h]
	\centering
	\begin{tikzpicture}
		\draw[->] (-2.5,0)--(4,0);
		\draw[->] (-1.5,0)--(-1.5,4);
		\draw[ thick,blue] (-1.5,0.54212)--(-1.5, 4);
		\draw[ thick,domain =-1.5:4.0,smooth,blue] plot(\x,{3/(cosh(\x)*cosh(\x))))});
		\coordinate[label=right:$ U(x) $](Ux) at (-1.5,4.0);
		\coordinate[label=right:$ x $](x) at (4,0);
		\coordinate[label=above:$ U_0 $](U0) at (0,3);
		\coordinate[label=below:$ b $](b) at (0,0);
		\node[below left] at (-1.5,0) {$0$};
	\end{tikzpicture}\caption{P\"oschl-Teller potential with a wall.}
    \label{Fig.4}
\end{figure}
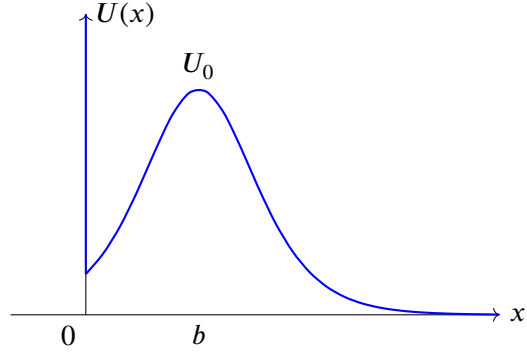

In this Appendix we explore the P\"oschl-Teller potential on the half line shown in figure~\ref{Fig.4}. The potential is,
\begin{equation}
	V(x)=
	\begin{cases}
		+\infty\,, & x<0\,,\\
		\frac{U_0}{\cosh^2[\alpha(x-b)]}\,,       & x\ge 0\,.\\
	\end{cases}
\end{equation}
\\
As with the P\"oschl-Teller potential on the whole line, we can write the eigenfunction~\cite{LANDAU197750},
\begin{align}
	\phi_k(x)=\frac{1}{N(k)}&\cosh^{-s}\!\bigl(\alpha (x-b)\bigr)\!
	\left\{
	C_1 \times{}_2F_1\!\left(-\frac12 s+\frac12 \frac{\i k}{\alpha\h}, -\frac12 s-\frac12 \frac{\i k}{\alpha\h}, \frac12, -\sinh^2\!\alpha (x-b)\right)\right.\notag\\
	&\hskip -1.4cm
    +\,
	\left.C_2\,\sinh(\alpha (x-b))\;{}_2F_1\!\left(-\frac12 s+\frac12 \frac{\i k}{\alpha\h}+\frac12, -\frac12 s-\frac12 \frac{\i k}{\alpha\h}+\frac12, \frac32, -\sinh^2\!\alpha (x-b)\right)\right\}
\,,
\end{align}
where 
\[
k = \sqrt{2mE}, 
\qquad
s = \tfrac{1}{2}\!\left(-1 + \sqrt{\,1 - \tfrac{8mU_{0}}{\h^2\alpha^{2}}}\,\right)
\,.\quad
\]
Using $\phi_k(0)=0$ we find,
\begin{equation}
	R(k)=\frac{C_1}{C_2}=\frac{\sinh(\alpha b)\;{}_2F_1\!\left(-\frac12 s+\frac12 \frac{\i k}{\alpha\h}+\frac12, -\frac12 s-\frac12 \frac{\i k}{\alpha\h}+\frac12, \frac32, -\sinh^2\!\alpha b\right)}{{}_2F_1\!\left(-\frac12 s+\frac12 \frac{\i k}{\alpha\h}, -\frac12 s-\frac12 \frac{\i k}{\alpha\h}, \frac12, -\sinh^2\!\alpha b\right)}
\,,\quad
\end{equation}
and hence we have,
\begin{align*}
	\phi_k(x)&=\frac{1}{N(k)}\cosh^{-s}\!\bigl(\alpha (x-b)\bigr)\!
	\left\{
	R(k)\times {}_2F_1\!\left(-\frac12 s+\frac12 \frac{\i k}{\alpha\h}, -\frac12 s-\frac12 \frac{\i k}{\alpha\h}, \frac12, -\sinh^2\!\alpha (x-b)\right)\right.\\
	&+
	\left.\,\sinh(\alpha (x-b))\times\;{}_2F_1\!\left(-\frac12 s+\frac12 \frac{\i k}{\alpha\h}+\frac12, -\frac12 s-\frac12 \frac{\i k}{\alpha\h}+\frac12, \frac32, -\sinh^2\!\alpha (x-b)\right)\right\}
\,.\quad
\end{align*}
%
Define a generalized Wronskian,
\begin{equation}
	W_{k,k'}(x)= \phi_k^*\,\partial_x\phi_{k'} - \partial_x\phi_k^*\,\phi_{k'}
\,.\quad
\end{equation}
For $-\h^2\phi''_k+2mV(x)\phi=k^2\phi_k$ and $-\h^2\phi_{k'}''+2mV\phi_{k'}=k'^2\phi_{k'}$, 
we have,
\begin{equation}
	\frac{(k^2-k'^2)}{\h^2}\int_{0}^{L}\phi_k^*(x)\,\phi_{k'}(x)\,\d x
	\;=\; \Big[W_{k,k'}(x)\Big]_{x=0}^{x=L}
\;.\quad
\label{eq:GL}
\end{equation}
With $\phi_k(0)=\phi_{k'}(0)=0\Rightarrow W_{k,k'}(0)=0$ we obtain,
\begin{equation}
	\frac{(k'^2-k^2)}{\h^2}\int_{0}^{L}\phi_k\,\phi_{k'}\,\d x \;=\; W_{k,k'}(L)
\,.\quad
\label{eq:GL-2}
\end{equation}
To determine the normalization factor $N(k)$, we can use the normalization condition for $x\to+\infty$, because the hypergeometric functions reduce to plane waves there,
\begin{equation}
	\lim_{x\to+\infty}\phi(x)=\frac{1}{N(k)}\left[A_+(k)\e^{\i kx/\h}+A_{-}(k)\e^{-\i kx/\h}\right]
\,,\quad
\end{equation}
where
\[ A_+=(A_2R+B_2)2^{-\i k/(\alpha\h)}\e^{-\i kb/\h}
\,,\qquad
A_-=(A_1R+B_1)2^{\i k/(\alpha\h)}\e^{\i kb/\h}, \]
and
\[
\hskip -0.5cm
A_1=\frac{\Gamma\!\left(\tfrac12\right)\Gamma\!\left(-\tfrac{\i k}{\alpha\h}\right)}
{\Gamma\!\left(-\tfrac12 s-\tfrac{\i k}{2\alpha\h}\right)\Gamma\!\left(\tfrac12+\tfrac12 s-\tfrac{\i k}{2\alpha\h}\right)}
\,,\quad
\qquad\quad
A_2=\frac{\Gamma\!\left(\tfrac12\right)\Gamma\!\left(\tfrac{\i k}{\alpha\h}\right)}
{\Gamma\!\left(-\tfrac12 s+\tfrac{\i k}{2\alpha\h}\right)\Gamma\!\left(\tfrac12+\tfrac12 s+\tfrac{\i k}{2\alpha\h}\right)}
\,,
\]
\[
\hskip 0.3cm
B_1=\frac{\Gamma\!\left(\tfrac32\right)\Gamma\!\left(-\tfrac{\i k}{\alpha\h}\right)}
{\Gamma\!\left(-\tfrac12 s-\tfrac{\i k}{2\alpha\h}+\tfrac12\right)\Gamma\!\left(1+\tfrac12 s-\tfrac{\i k}{2\alpha\h}\right)}
\,,
\qquad
B_2=\frac{\Gamma\!\left(\tfrac32\right)\Gamma\!\left(\tfrac{\i k}{\alpha\h}\right)}
{\Gamma\!\left(-\tfrac12 s+\tfrac{\i k}{2\alpha\h}+\tfrac12\right)\Gamma\!\left(1+\tfrac12 s+\tfrac{\i k}{2\alpha\h}\right)}
\,.\quad
\]
Here we have used the identity,
\begin{align}
	{}_2F_1(\alpha,\beta,\gamma,z)&=
	\frac{\Gamma(\gamma)\Gamma(\beta-\alpha)}{\Gamma(\beta)\Gamma(\gamma-\alpha)}
	(-z)^{-\alpha}\,
	{}_2F_1\!\left(\alpha,\alpha+1-\gamma,\alpha+1-\beta,\frac{1}{z}\right)\\
	&+\frac{{\Gamma(\gamma)\Gamma(\alpha-\beta)}}{{\Gamma(\alpha)\Gamma(\gamma-\beta)}}
	(-z)^{-\beta}\,
	{}_2F_1\!\left(\beta,\beta+1-\gamma,\beta+1-\alpha,\frac{1}{z}\right)
\,.\quad
\end{align}
For the wave function we have the Wronskian identity,
\begin{equation}
	W=\phi_k^*\phi_k'-\phi_k(\phi_k')^*=\frac{2\i k}{\h}\bigl(|A_+(k)|^2-|A_-(k)|^2\bigr)=\phi_k(0)^*\phi_k(0)'-\phi_k(0)\bigl[\phi_k(0)'\bigr]^*=0
\,,\quad
\label{Wronskian identity}
\end{equation}
and hence,
\[|A_+(k)|=|A_-(k)|
\,.\quad
\]
Next, setting $A_\pm(k)=|A(k)|\e^{\i\theta_\pm}$ 
we have,
\begin{align*}
	\lim_{x\to+\infty}\phi_k(x)
	&=\frac{|A(k)|}{N(k)}\left[\e^{\i(\theta_++kx/\h)}+\e^{\i(\theta_--kx/\h)}\right]\\
	&=\frac{2|A(k)|}{N(k)}\e^{\frac{\i}{2}(\theta_++\theta_-)}\cos\left(kx+\frac{1}{2}(\theta_+-\theta_-)\right)\\
	&=\frac{2|A(k)|}{\widetilde{N}(k)}\cos\left(kx+\frac{1}{2}(\theta_+-\theta_-)\right)
\,,\quad
\end{align*}
where $|\widetilde{N}(k)|=|N(k)|$. Let $\theta_k=kx/\h+\delta_k$, then
\begin{align}
	\lim_{x\to+\infty}W_{k,k'}(x)
	&= \phi_k^*\,\partial_x\phi_{k'} - \partial_x\phi_k^*\,\phi_{k'}\notag\\
	&= \frac{4|A|^2}{\widetilde{N}(k)\widetilde{N}^*(k')}\Big[-k'/\h\cos\theta_k\,\sin\theta_{k'} + k/\h\sin\theta_k\,\cos\theta_{k'}\Big] \notag\\
	&= \frac{2|A|^2}{\widetilde{N}(k)\widetilde{N}^*(k')}
	\Big[(k{-}k')/\h\sin\!\big(\theta_k{+}\theta_{k'}\big) + (k{+}k')/\h\sin\!\big(\theta_k{-}\theta_{k'}\big)\Big]
\,.\quad
\label{eq:Wronskian-cos}
\end{align}
Insert \eqref{eq:Wronskian-cos}  to \eqref{eq:GL-2} and $\theta_k=kL/\h+\delta_k$, $\theta_{k'}=k'L/\h+\delta_{k'}$, for large $L$ we have,
\begin{align}
	\int_{0}^{L}\phi^*_k\,\phi_{k'}\,\d x
	&=  \frac{2|A|^2\h}{\widetilde{N}(k)\widetilde{N}^*(k')}
	\left[
	\frac{\sin\!\big((k{-}k')L/\h+\delta_k{-}\delta_{k'}\big)}{k{-}k'}
	+ \frac{\sin\!\big((k{+}k')L/\h+\delta_k{+}\delta_{k'}\big)}{k'{+}k}
	\right]
\,.\qquad
\end{align}
As $L\to\infty$, using the distributional identity $\lim_{L\to+\infty}\sin(Lx)/(\pi x)=\delta(x)$, we have,
\begin{equation}	\lim_{L\to+\infty}\frac{\sin\!\big((k{-}k')L/\h+\Delta\big)}{(k{-}k')/\h}=\pi\,\cos(\Delta)\delta(k-k')=\pi\,\delta(k-k')
\,.\quad
\end{equation}
\begin{equation}	\lim_{L\to+\infty}\frac{\sin\!\big((k{+}k')L/\h+\Sigma\big)}{(k'{+}k)/\h}=-\pi\cos(\Sigma)\delta(k+k')=0
\,,\quad
\end{equation}
with finite phases $\Delta=\delta_k-\delta_{k'}$, $\Sigma=\delta_k+\delta_{k'}$.
Therefore,
\begin{equation}
	\int_{0}^{\infty}\phi_k(x)\,\phi_{k'}(x)\,\d x
	= \frac{2\pi|A|^2}{|N(k)|^2}\,\pi\,\delta(k-k')
\,.\quad
\label{eq:inner-product}
\end{equation}
Impose the normalization condition,
\begin{equation}
	\int_{0}^{\infty}\phi^*_k(x)\,\phi_{k'}(x)\,\d x \;=\; \delta(k-k')
\,,\quad
\end{equation}
and read off from \eqref{eq:inner-product} that,
\begin{equation}
	|N(k)|=\sqrt{2\pi}\,|A(k)|
	\;=\;\sqrt{2\pi}\,|A_-(k)|
	\;=\;\sqrt{2\pi}\,|A_+(k)|
\,,\quad
\end{equation}
and hence,
\begin{equation}
	|N(k)|^2=2\pi |(A_1(k)R(k)+B_1(k))|^2=2\pi|(A_2(k)R(k)+B_2(k))|^2
\,.\quad
\end{equation}

\subsection{Calculating the kernel}
To calculate the kernel, we must first establish the semi-classical condition. The de Broglie wavelength under the barrier is $\lambda\sim\h/\sqrt{2m(U_0-E)}=\h/\kappa$, and the potential changes over the length $\Delta x\sim/\alpha$, and hence,
\[ \Delta x\gg \lambda\implies\frac{\kappa}{\h\alpha}\gg 1
\,.\quad
\]
The kernel is,
\begin{equation}
	K(t,x;0,y)=\int_{0}^{\infty} \frac{\d E}{2\pi\h}\sqrt{\frac{m}{2E}} \,
	\phi_{k}^{*}(y)\,\phi_{k}(x)\,
	\e^{-\i Et/\h}
\,,\quad
\end{equation}
for $x,y\in\mathrm{L}$. According to the definition of the left region, we assume the initial energy of the particle $E\ll U_0$, therefore $x,y\ll b$. Notice that,
\begin{equation}
	\lim_{x\ll b}\phi(x)=\frac{1}{N(k)}\left[(A_1R-B_1)2^{\frac{\i k}{\alpha\h}}\e^{\i k(x-b)/\h}+(A_2R-B_2)2^{-\frac{\i k}{\alpha\h}}\e^{-\i k(x-b)/\h}\right]
\,,\quad
\end{equation}
and hence,
\begin{equation}
	K(t,x;0,y)=\int_{0}^{\infty} \frac{\d E}{2\pi\h}\sqrt{\frac{m}{2E}} \,
	\frac{1}{|N(k)|^2}\left[|A_1R-B_1|^2\e^{\i k(x-y)/\h}+|A_2R-B_2|^2\e^{-\i k(x-y)/\h} \right]
	\e^{-\i Et/\h}
\,.\quad
\end{equation}
Similarly to~\eqref{Wronskian identity}, 
we can prove that,
\[ |A_2R-B_2|=|A_1R-B_1|
\,,\quad
\]
and then we find,
\begin{align}
	K(t,x;0,y)&=\int_{0}^{\infty} \frac{\d E}{2\pi\h}\sqrt{\frac{m}{2E}} \,\frac{|A_2R-B_2|^2}{|N(k)|^2}\left(\e^{\i k(x-y)/\h}+\e^{-\i k(x-y)/\h}\right)\e^{-\i Et/\h}\notag\\
	&=\int_{0}^{\infty} \frac{\d E}{2\pi\h}\sqrt{\frac{m}{2E}} \,\frac{|A_2R-B_2|^2}{|N(k)|^2}2\cos(k(x-y)/\h)\e^{-\i Et/\h}
\,.\quad
\end{align}
We now analytically continue $|N(E)|^2$ and $|A_2 R-B_2|^2$ to a domain of complex energies and deform the contour to get,
\begin{equation}
	K(t,x;0,y)=f(t,x,y)+g(t,x,y)\,,
\end{equation}
where
\begin{equation}
	f(t,x,y)=\oint_{\mathcal{C}} \frac{\d E}{2\pi\h}\sqrt{\frac{m}{2E}} \,\frac{|A_2R-B_2|^2}{|N(k)|^2}2\cos(k(x-y)/\h)\e^{-\i Et/\h}
\,,\quad
\end{equation}
and
\begin{align}
	g(t,x,y) &= \int_{0}^{-\i\infty} \frac{\d E}{2\pi\h}\sqrt{\frac{m}{2E}} \,\frac{|A_2R-B_2|^2}{|N|^2}2\cos(\sqrt{2mE}(x-y)/\h)\e^{-\i Et/\h}\\
    &=-\i\int_{0}^{\infty} \frac{\d \mathcal{E}}{2\pi\h}\sqrt{\frac{m}{-2\i\mathcal{E}}} \,\frac{|A_2R-B_2|^2}{|N|^2}2\cos(\sqrt{-2m\i\mathcal{E}}(x-y)/\h)\e^{-\mathcal{E}t/\h}
\,.\quad
\end{align}
The normalization factor $|N(k)|^2$ has two families of zeros: one located on the imaginary axis, and the other located in the lower half-plane close to the real axis. The former has no physical meaning, and we evaluate $f(t,x,y)$ using the latter, $E_n\!-\!\i b_n$. 
We thus have,
\begin{align}
	f(t,x,y) 
	&= -\frac{\i}{2\pi} \sum_{n} 
	\sqrt{\frac{m}{2E_n}}\,
	\frac{|A_2R-B_2|^2}{|N'(k)N^*(k)|}2\cos(k(x-y)/\h) \,\e^{-\i E_n t/\h-b_n t/\h}\notag\\[8pt]
	&\approx -\frac{\i}{2\pi}\sqrt{\frac{m}{2E_0}}\,
	\frac{|A_2R-B_2|^2}{|N'(k)N^*(k)|}2\cos(k(x-y)/\h) \,\e^{-\i E_0 t/\h-b_0 t/\h}
\,,\quad
\end{align}
where the last approximation holds for $t \gg \h/\Gamma_0$ (late times). 

To evaluate $g(t,x,y)$, Setting $\mathcal{E} \to E_0 \alpha$,
\begin{equation}
    g(t,x,y)=-\i \frac{E_0}{2\pi\h}\sqrt{\frac{m}{-2\i E_0}}\int_{0}^{\infty} \d\alpha\ \frac{1}{\sqrt{\alpha}}\,\frac{|A_2R-B_2|^2}{|N|^2}2\cos(\sqrt{-2m\i E_0 \alpha}(x-y)/\h)\e^{- E_0 \alpha t/\h}
\,.\quad
\end{equation}
In the late time limit, only small $\alpha$ contribute, and hence,
\begin{align}
g(t,x,y)
&= -\i \frac{E_{0}}{2\pi\h}\sqrt{\frac{m}{-2\i E_{0}}}\frac{|A_2(0)R(0)\!-\!B_2(0)|^2}{|N(0)|^2}
\int_{0}^{\infty} \!\! \d\alpha\,\alpha^{-1/2}\,
2\cos\!\left(\sqrt{-2m\i E_{0}\alpha}\,\frac{x\!-\!y}{\h}\right)
\e^{-E_{0}\alpha t/\h}
\\[4pt]
&=\sqrt{\frac{m}{2\pi\h}}\frac{|A_2(0)R(0)\!-\!B_2(0)|^2}{|N(0)|^2}\,
\e^{-\i\pi/4}\,t^{-1/2}\,
\exp\!\left[\frac{\i m (x\!-\!y)^{2}}{2\h t}\right]
\\[4pt]
&= \frac{1}{\h}\,\sqrt{\frac{mE_0}{2\pi}}\frac{|A_2(0)R(0)\!-\!B_2(0)|^2}{|N(0)|^2}\,
\e^{-\i\pi/4}
(E_0t/\h)^{-1/2}+\mathcal{O}\!\bigl(t^{-3/2}\bigr)
\,.\quad
\end{align}
Therefore, the kernel scales as,
\begin{align}
K(t,x;0,y) &\approx
	G(E_0,x,y)\,(E_0 t/\h)^{-1/2}
	+  \sum_{n} F(E_n,x,y)\,\exp\!\left\{-\tfrac{\Gamma_n}{2} t\right\} \notag \\[6pt]
	&\approx G(E_0,x,y)\,(E_0 t/\h)^{-1/2}
	+ F(E_0,x,y)\,\exp\!\left\{-\tfrac{\Gamma_0}{2} t\right\}
\,.\quad
\end{align}

\subsection{Explicit calculation of the rate}\label{P\"oschl-Teller potential explicit}

Here we perform an explicit calculation of the rate
and twilight time. For convenience we assume $b\to+\infty$. We need to find the zeros of $A_+(k)=A_2(k)R(k)+B_2(k)=0$. Notice that,
\begin{equation}
	\lim_{b\to+\infty}R(k)=\frac{ B_1\,\e^{-\i k b}\,2^{\,\frac{\i k}{\alpha}} + B_2\,\e^{\i k b}\,2^{-\frac{-\i k}{\alpha}} }
	{ A_1\,\e^{-\i k b}\,2^{\,\frac{\i k}{\alpha}} + A_2\,\e^{\i k b}\,2^{-\frac{\i k}{\alpha}} }
\,,\quad
\end{equation}
and hence we have,
\begin{equation}
	(A_2B_1+B_2A_1)2^{\i k/\alpha}\e^{-i kb}+2A_2B_2 2^{-\i k/\alpha}\e^{\i kb}=0\label{k_0C}
\,,\quad
\end{equation}
\begin{equation}
	\e^{2\i k L}=-\frac{1}{2}\left[\frac{A_1}{A_2}+\frac{B_1}{B_2}\right]
\,,\quad
\end{equation}
where $L=b-\frac{\ln 2}{\alpha}$. In the thick barrier limit $\frac{\sqrt{2m(U_0-E)}}{\h\alpha}\gg 1$, $|\frac{A_1}{A_2}|=|\frac{B_1}{B_2}|=1$, and hence we have approximate real roots $k_{R,n}=\frac{(2n+1)\pi}{2L}$. It is difficult to solve the complex solutions and calculate the decay rate $\Gamma$ directly, but we can deploy an "inverse" approach. Just like what we have discussed in Appendix~\ref{suqre barrier} and~\ref{Modified square barrier}, for large $b\to+\infty$ and the thick barrier condition $\kappa/{(\h\alpha)}\gg 1$ we then have, 
\begin{equation}
    T_{\rm trans}=T\Gamma
\,,\quad
\end{equation}
where $T=\frac{2mb}{k_R}$ is the oscillation period and $T_{\rm trans}$ is the transmission coefficient for P\"oschl-Teller potential,
\[
T_{\rm trans}=\frac{\sinh^{2}\!\left(\tfrac{\pi k}{\alpha\h}\right)}{\sinh^{2}\!\left(\tfrac{\pi k}{\alpha\h}\right) + \cosh^{2}\!\left( \tfrac{\pi}{2}\sqrt{\tfrac{8mU_{0}}{\alpha^{2}\h^2}-1} \right)}
\,.\quad
\]
Hence we could find 
\begin{align}
    \Gamma=\lim_{\kappa/{(\h\alpha)}\gg 1}
    \frac{T_{\rm trans}}{T}=\frac{2k_R}{bm} \sinh^2(\frac{\pi k_R}{\alpha\h})\e^{-\frac{2\pi\kappa}{\h\alpha}}
\,.\quad
\label{Gamma for PT}    
\end{align}
Here we could also calculate the twilight time using \eqref{Gamma for PT}, 
\begin{equation}
	t_\circletophalfblack^{(C)} \;=\;
	-\,\frac{1}{\Gamma_0}\;
	W_{-1}\!\left(
	-\,\frac{\Gamma_0\h}{E_0}\;
	\left|
\frac{G}{F}
	\right|^2
	\right)
\,.\quad
\end{equation}
Using $W_{-1}(-\epsilon)=\ln(\epsilon)+\mathcal{O}(\ln|\ln\epsilon|)$ for $\epsilon\to0$, we finally get,
\begin{equation}	t_\circletophalfblack^{(C)}\approx\frac{2\pi}{\Gamma_0}\frac{\kappa}{\h\alpha}
\,,\qquad
\end{equation}
which is Eq.~(\ref{twilight time: estimate 2}) in the main text.

\nocite{*}
		\bibliography{ref}
		
	\end{document}